\documentclass[]{aa}
\usepackage[varg]{txfonts}
\usepackage{graphicx}
\usepackage[normalem]{ulem}
\usepackage{aecompl,color, amsmath}
\usepackage{siunitx}
\DeclareSIUnit\erg{erg}
\graphicspath{ {figures/} }

\begin{document}

\title{Studying the ISM at $\sim$ 10 pc scale in NGC 7793 with MUSE}
\subtitle{II. Constraints on the oxygen abundance and ionising radiation escape}
\author{
Lorenza Della Bruna\inst{\ref{inst1}} \and
Angela Adamo\inst{\ref{inst1}} \and
Janice~C. Lee\inst{\ref{inst2}} \and
Linda~J. Smith\inst{\ref{inst3}} \and 
Mark Krumholz\inst{\ref{inst4}} \and
Arjan Bik\inst{\ref{inst1}} \and
Daniela Calzetti\inst{\ref{inst5}} \and
Anne Fox\inst{\ref{inst1}}  \and
Michele Fumagalli\inst{\ref{inst6}, \ref{inst7}, \ref{inst8}}  \and
Kathryn Grasha\inst{\ref{inst4}} \and
Matteo Messa\inst{\ref{inst5}, \ref{inst1}} \and
Göran Östlin\inst{\ref{inst1}}  \and
Rene Walterbos\inst{\ref{inst9}} \and
Aida Wofford\inst{\ref{inst10}}
}

\institute{
Department of Astronomy, Oskar Klein Centre, Stockholm University, AlbaNova University
Centre, SE-106 91 Stockholm, Sweden\label{inst1}
\and
Infrared Processing and Analysis Center, California Institute of Technology, Pasadena, CA\label{inst2}
\and
European Space Agency (ESA), ESA Office, Space Telescope  Science Institute, 3700 San Martin Drive, Baltimore, MD 21218, USA\label{inst3}
\and
Research School of Astronomy and Astrophysics, Australian National University, Canberra, ACT 2612, Australia\label{inst4}
\and
Observatoire de Genève, Université de Genève, Chemin Pegasi 51, 1290 Versoix, Switzerland\label{inst5}
\and
Dipartimento di Fisica G. Occhialini, Universit\`a degli Studi di Milano Bicocca, Piazza della Scienza 3, 20126 Milano, Italy\label{inst6}
\and
Institute for Computational Cosmology, Durham University, South Road, Durham, DH1 3LE, UK\label{inst7}
\and
Centre for Extragalactic Astronomy, Durham University, South Road, Durham, DH1 3LE, UK\label{inst8}
\and
Department of Astronomy, New Mexico State University, Las Cruces, NM, 88001, USA\label{inst9}
\and
Universidad Nacional Autónoma de México, Instituto de Astronomía, AP 106,  Ensenada 22800, BC, México\label{inst10}
}

\authorrunning{Della Bruna et al.}
\titlerunning{Studying the ISM at $\sim$ 10 pc scale in NGC 7793}

\date{Received 11 September 2020 / Accepted 15 April 2021}

 \abstract
%%% * Context
{Feedback from massive stars affects the interstellar medium (ISM) from the immediate surroundings of the stars (parsec scales) to galactic (kiloparsec) scales. High-spatial resolution studies of \ion{H}{ii} regions are critical to investigate how this mechanism operates.
}
%%% * Aims
{
We study the ionised ISM in NGC 7793 with the MUSE instrument at ESO Very Large Telescope (VLT), over a field of view (FoV) of $\sim$ 2 kpc$^2$ and at a spatial resolution of $\sim$ 10 pc.
The aim is to link the physical conditions of the ionised gas (reddening, ionisation status, abundance measurements) within the spatially resolved \ion{H}{ii} regions to the properties of the stellar populations producing Lyman continuum photons.
}
%%% * Methods
{The analysis of the MUSE dataset, which provides a map of the ionised gas and a census of Wolf Rayet stars, is
complemented with a sample of young star clusters (YSCs) and O star candidates observed with the Hubble Space Telescope (HST) and of giant molecular clouds traced in CO(2-1) emission with the Atacama Large Millimeter/submillimeter Array (ALMA).
We estimated the oxygen abundance using a temperature-independent strong-line method. We determined the observed total amount of ionising photons ($Q(H^0)$) from the extinction corrected H$\alpha$ luminosity. This estimate was then compared to the expected $Q(H^0)$ obtained by summing the contributions of YSCs and massive stars. The ratio of the two values gives an estimate for the escape fraction ($f_{esc}$) of photons in the region of interest. We used the [\ion{S}{ii}]/[\ion{O}{iii}] ratio as a proxy for the optical depth of the gas and classified \ion{H}{ii} regions into ionisation bounded, or as featuring channels of optically thin gas. We compared the resulting ionisation structure with the computed $f_{esc}$. We also investigated the dependence of $f_{esc}$ on the age spanned by the stellar population in each region.
}
%%% * Results
{
We find a median oxygen abundance of $12 + \log(O/H) \sim 8.37$, with a scatter of 0.25 dex, which is in agreement with previous estimates for our target. 
We furthermore observe that the abundance map of \ion{H}{ii} regions is rich in substructures, surrounding clusters and massive stars, although clear degeneracies with photoionisation are also observed.
From the population synthesis analysis, we find that YSCs located in \ion{H}{ii} regions have a higher probability of being younger and less massive as well as of emitting a higher number of ionising photons than clusters in the rest of the field.
Overall, we find $f_{esc, \ion{H}{ii}} = 0.67_{-0.12}^{+0.08}$ for the population of \ion{H}{ii} regions.
We also conclude that the sources of ionisation observed within the FoV are more than sufficient to explain the amount of diffuse ionised gas (DIG) observed in this region of the galaxy.
We do not observe a systematic trend between the visual appearance of \ion{H}{ii} regions and $f_{esc}$, pointing to the effect of 3D geometry in the small sample probed.}
{} 
\maketitle

\section{Introduction}
\label{section:introduction}
As a result of hierarchically-structured, accreting molecular clouds, stars preferentially form in clustered environments. More than 70 \% of massive stars are located in clusters and OB associations \citep{ladalada03}, making young star clusters (YSCs) preferential sites from which radiative and mechanical feedback originates. Typically, only a minority of these stellar aggregates are gravitationally bound and will, therefore, survive as star clusters \citep{kruijssen12a, grudic20}. These are the clusters that sit in the densest regions of the hierarchy and are therefore able to survive the removal of gas long enough to reach a high star formation efficiency and become dynamically relaxed and well-mixed \citep{kruijssen19}. The remaining loosely-bound clusters will disperse in short timescales and blend into the field stellar population.

\par A detailed understanding of the process of stellar feedback is important as this mechanism leads to the suppression of global star formation in galaxies and to a multiphase and metal-enriched interstellar medium (ISM).
In particular, it will shed light on the origin of the diffuse ionised gas (DIG) that has been observed to constitute up to 50\% of the total H$\alpha$ luminosity in local spiral galaxies \citep{ferguson96, hoopes96, zurita00, thilker02, oey07}. The origin of this gas is unclear \citep[see e.g.][for a review]{mathis00, haffner09}, and it has been linked to various processes including leaking \ion{H}{ii} regions~\citep{zurita02, weilbacher18}, evolved field stars \citep{hoopes00,zhang17}, shocks~\citep{collins01}, and cosmic rays~\citep{vandenbroucke18}.
Both theory and observation have shown that stellar feedback can facilitate the escape of ionising radiation from the star forming regions \citep[see e.g.][]{bik15,menacho19} if the youngest star clusters have had time
to clear out channels through the dense leftover gas from the parent giant molecular clouds (GMCs) from which they were born~\citep{dale15,howard18}.
In order to study how the radiation from massive stars affects the ISM, it is therefore necessary to investigate the timescale for photons to find a channel through which they can leave the  region.
Various studies have found that small scale physics plays an important role in the escape of ionising radiation. In particular, the effect of temporal and spatial clustering of supernovae (SN) can have important consequences \citep{gentry17,kim17,fielding18}, resulting in a transferred momentum up to an order of magnitude larger than for isolated SN.
Simulations of individual galaxies do indeed indicate that the escape of ionising radiation is largely driven by the distribution of dense gas \citep[see e.g.][]{paardekooper11}, and high-resolution GMC simulations have proven that photons are more likely to escape from less massive GMCs \citep{howard18}. Consequently, late-emitted ionising photons are more likely to escape the star forming region as the gas surrounding the region is dispersed by ionising radiation and supernova feedback~\citep{kim13a}. 
The same trend has been recently observed in high-resolution cosmological simulations, showing that photons are preferentially leaking from star forming regions containing feedback-driven superbubbles~\citep{ma20}.
Finally, stars are also responsible of yielding metals that will enrich the ISM, and a detailed understanding of abundance variations in star forming regions is needed in order to constrain chemical evolution models. However, until recently, it has been unclear whether star forming galaxies feature, in general, a homogeneous metal enrichment or whether the enrichment varies in different star forming regions and within the regions themselves.
Recent high-resolution studies \citep[e.g.][]{james16, mcleod19} seem to indicate the latter, implying that feedback mechanisms might play an important role in local metallicity variations.

\par High resolution studies of star forming regions are therefore an essential piece of information in order to bridge the spatial scales that critically connect the sources of feedback (stars and clusters, i.e. a few parsec scales) and their immediate surroundings (10s parsec scale) to galactic scale dynamics (kpc scale).
Such studies, initially limited to narrow-band photometry \citep[e.g.][]{pellegrini12}, were revolutionised with the advent of instruments with high spatial resolution integral field spectroscopy capabilities, such as the Multi Unit Spectroscopic Explorer~\citep[MUSE,][]{bacon10} instrument at ESO Very Large Telescope (VLT). Recently, \citet{mcleod19,mcleod20} exploited the MUSE capability for a detailed study of \ion{H}{ii} regions in the Large Magellanic Cloud and in the dwarf spiral NGC 300, mapping the ionisation structure of the regions and constraining their escape of ionising radiation.
In this work, we perform a similar study in the local spiral galaxy NGC 7793 at a spatial resolution of $\sim$ 10 pc with MUSE. In \citet[][henceforth Paper~\textsc{i}]{paperI} we presented the data, constructed a  sample of \ion{H}{ii} regions, determined the fraction of DIG in our field of view (FoV) and studied the properties of the ionised gas in emission line diagrams.
Here, we combine the MUSE results on the ionised gas with the stellar and cluster population studied with the Hubble Space Telescope (HST) and the molecular gas mapped with the Atacama Large Millimeter/submillimeter Array (ALMA),
construct a photoionisation budget for each \ion{H}{ii} region and study how it relates to its stellar and GMC content and to its ionisation structure.

% * Structure of the paper 
\par This work is structured as follows.
In Sect.~\ref{section:data} we give a brief overview of the data.
In Sect.~\ref{section:stellar_pop} we present the stellar census in our FoV and analyse the distribution of YSC properties.
In Sect.~\ref{section:extinction} we compare extinction maps derived from individual stellar extinction and from the ionised gas and in Sect.~\ref{section:metallicity} we derive the oxygen abundance of the \ion{H}{ii} regions.
In Sect.~\ref{section:optical_depth} we analyse the ionisation structure of the gas, and in Sect.~\ref{section:budget} we compile a photoionisation budget for the entire FoV and in more detail for a subset of the \ion{H}{ii} regions.
Finally, in Sect.~\ref{section:discussion} we discuss our findings and in Sect.~\ref{section:conclusions} we give a brief summary and present our conclusions.

\begin{figure*}
\center
\includegraphics[width=16cm]{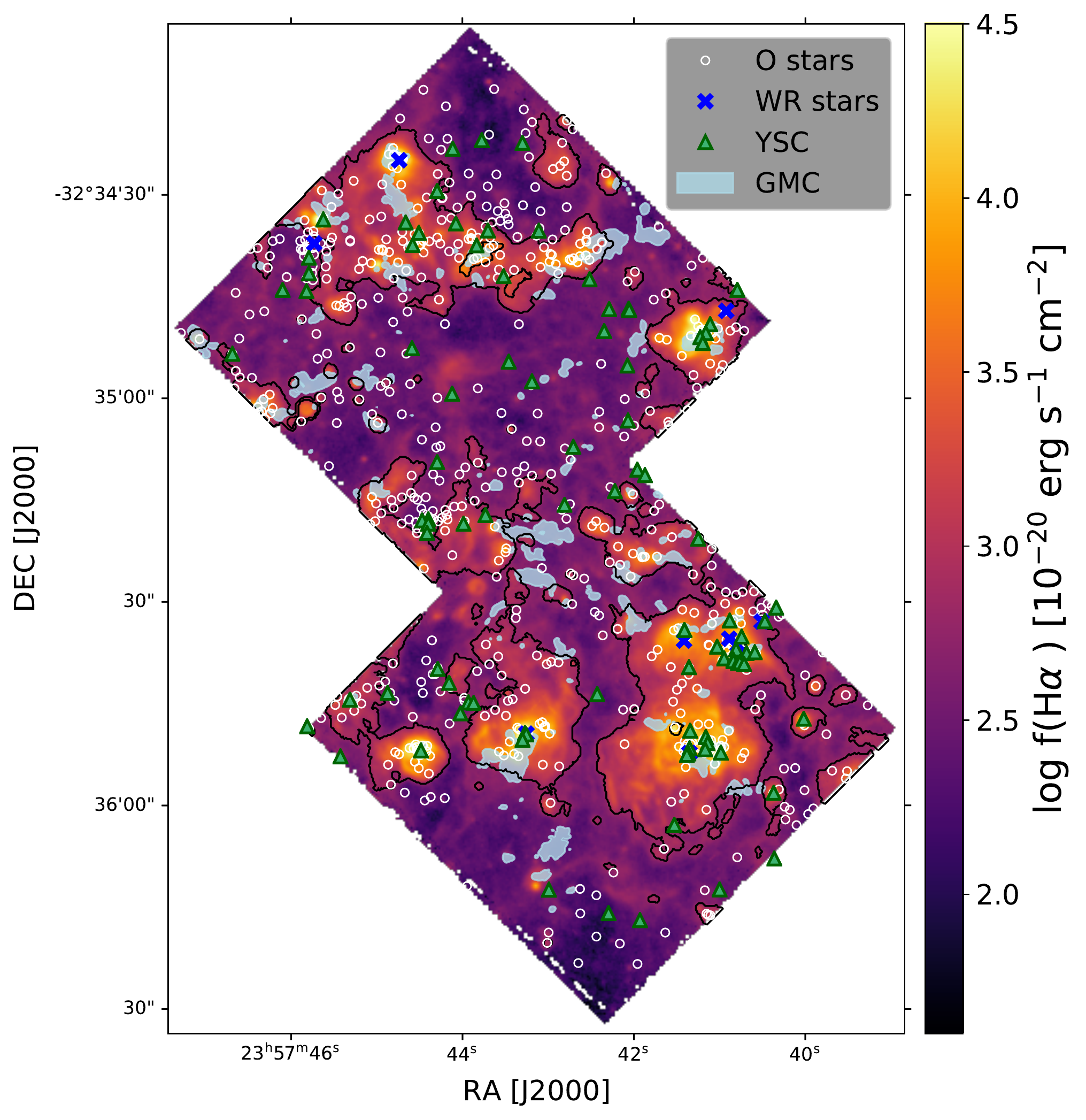}
\caption{Stellar census in the MUSE FoV. The location of candidate O stars (white circles), YSCs (green triangles) and WR stars (blue crosses) is indicated, overlaid on a map of the H$\alpha$ emission. Black contours indicate the \ion{H}{ii} regions identified in Paper~\textsc{i}, light blue filled contours denote the position of GMCs.}
\label{fig:stellar_census}
\end{figure*}

\section{Data description}
\label{section:data}
The data and data reduction steps are described in detail in Paper~\textsc{i}: we give here a brief summary. We observed the nearby flocculent spiral galaxy NGC 7793 with MUSE\footnote{ESO programme 60.A-9188(A), PI Adamo.}.
The dataset consists of two pointings in the Wide Field Mode (WFM) adaptive optics assisted (AO) configuration, with the extended wavelength setting. On the final datacube, we measure a PSF of $0.62 \arcsec$ FWHM (at 7000~\AA{}), corresponding to a spatial resolution of $\sim$ 10 pc at the distance of NGC 7793.
\par NGC 7793 has also been observed in the framework of the HST Treasury programme LEGUS\footnote{HST GO–13364.}~\citep{calzetti15} and in ALMA CO(J = $2-1$).
% * YSCs
The catalogues of YSCs derived from the LEGUS data are publicly available\footnote{\url{https://archive.stsci.edu/prepds/legus/dataproducts-public.html}}: in this work, we make use of the catalogue generated with the Geneva stellar evolution models and Milky Way extinction. We are moreover selecting only clusters labelled as `Class 1' (compact and symmetric), `Class 2' (concentrated but with some degree of asymmetry), and `Class 3' (multiple peak systems). For more details on the cluster catalogues (e.g. cluster identification, classification and photometry) see~\citet{adamo17}. 
% * O stars
Lee et al. (in preparation) have identified O stars based on HST photometry. The catalogue is described in \citet{wofford20}, and consists of
main sequence O stars of $M > 20~M_\odot$, selected in a colour-magnitude and colour-$Q$ diagram\footnote{Here, $Q$ is the reddening-free parameter introduced by \citet{johnson53}.}.
The resulting completeness of this stellar catalogue is estimated to be about 75\% at the lower mass limit of $20~M_\odot$ and rapidly growing to 90\% at stellar masses above $25~M_\odot$\footnote{We note that the mass of the candidates is not constrained by the selection method. However, mass estimates have been obtained from completeness simulations, in which the candidates are compared to synthetic stars.}. The authors report also an average contamination rate $\sim$ 30\%. We use the latter rate to account for uncertainties in the contribution of O stars to the ionisation budget.

% * GMCs
The catalogue of GMCs built on the ALMA dataset is presented in \citet{grasha18}. The ALMA data have an angular resolution of $0.85''$ and a velocity resolution of 1.2 km s$^{-1}$. The data are sensitive to emission up to $11''$ and are mapping GMC down to masses $\sim 10^4 M_{\sun}$.

\section{Properties of the stellar population}
\label{section:stellar_pop}

\begin{figure*}%[ht]
\center
\includegraphics[width=18cm]{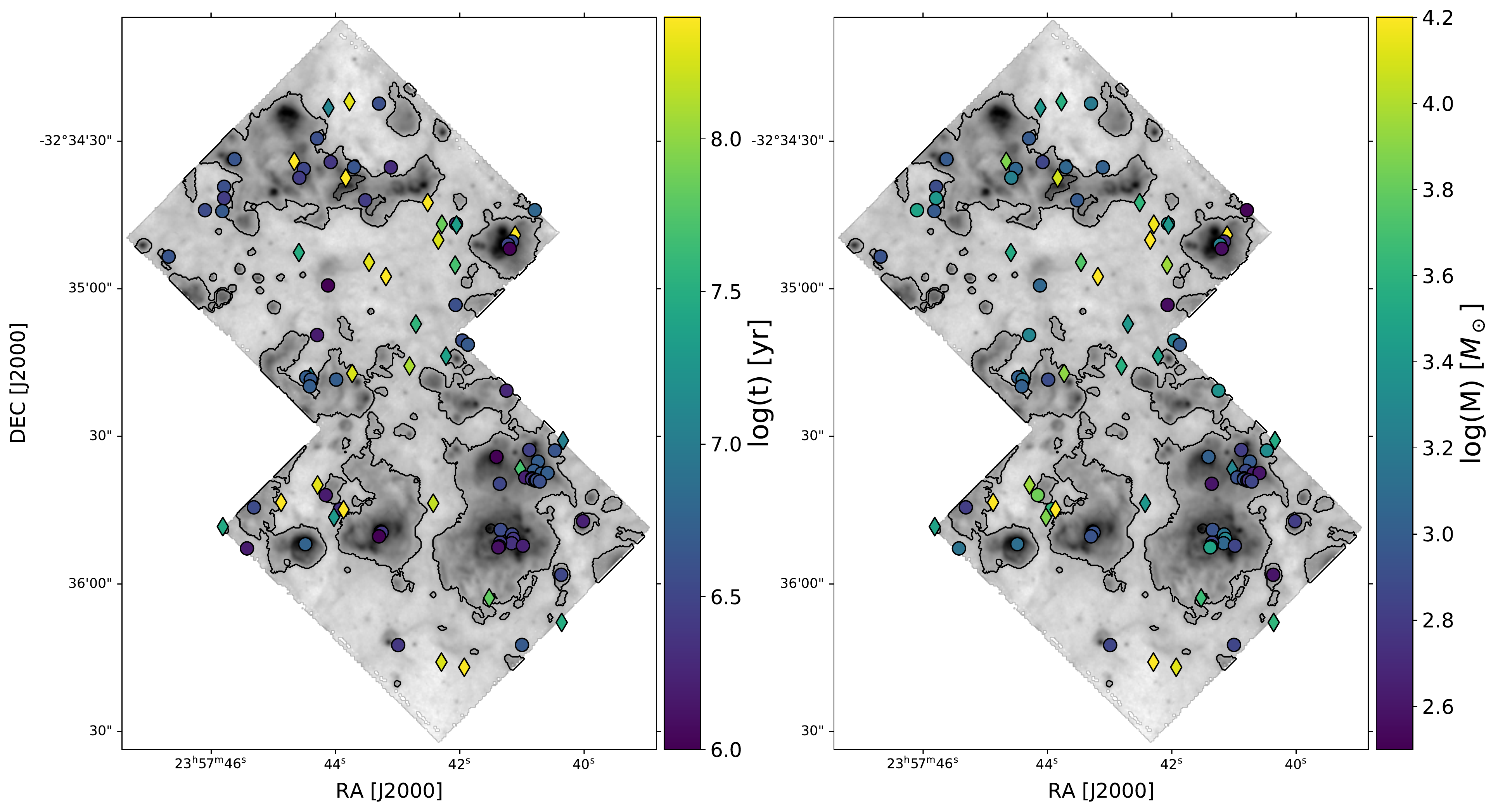}
\caption{Spatial distribution of YSC ages (\textit{left panel}) and masses (\textit{right panel}) in the MUSE FoV, overlaid on the H$\alpha$ map shown in Fig.~\ref{fig:stellar_census}. Circles and diamonds indicate, respectively, clusters younger and older than 10 Myr.
Ages and masses are best estimates computed from the median of the corresponding PDF (see Sect.~\ref{section:ysc_properties}).
Black contours indicate the \ion{H}{ii} regions identified in Paper~\textsc{i}.}
\label{fig:ysc_distr_FOV}
\end{figure*}

\begin{figure*}%[ht]
\center
\includegraphics[width=17cm]{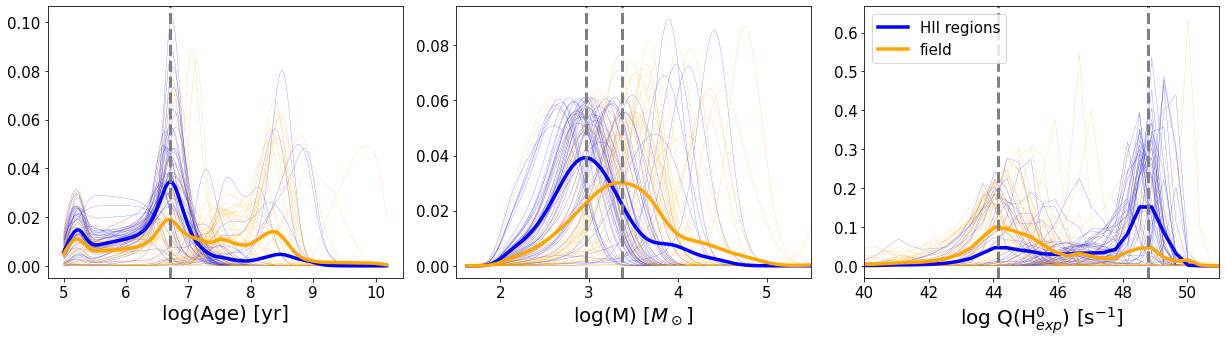}
\caption{Posterior probability distributions of YSC age (\textit{left}) mass (\textit{centre}) and ionising flux (\textit{right}) for clusters located in \ion{H}{ii} regions (in blue) and for field clusters (in orange). Thick lines indicate the combined PDFs, thin lines correspond to the PDFs of single clusters. The grey dashed lines indicate the location of the main modes of the combined distributions.}
\label{fig:ysc_agemass_hist}
\end{figure*}

\begin{figure}%[ht]
\center
\includegraphics[width=7cm]{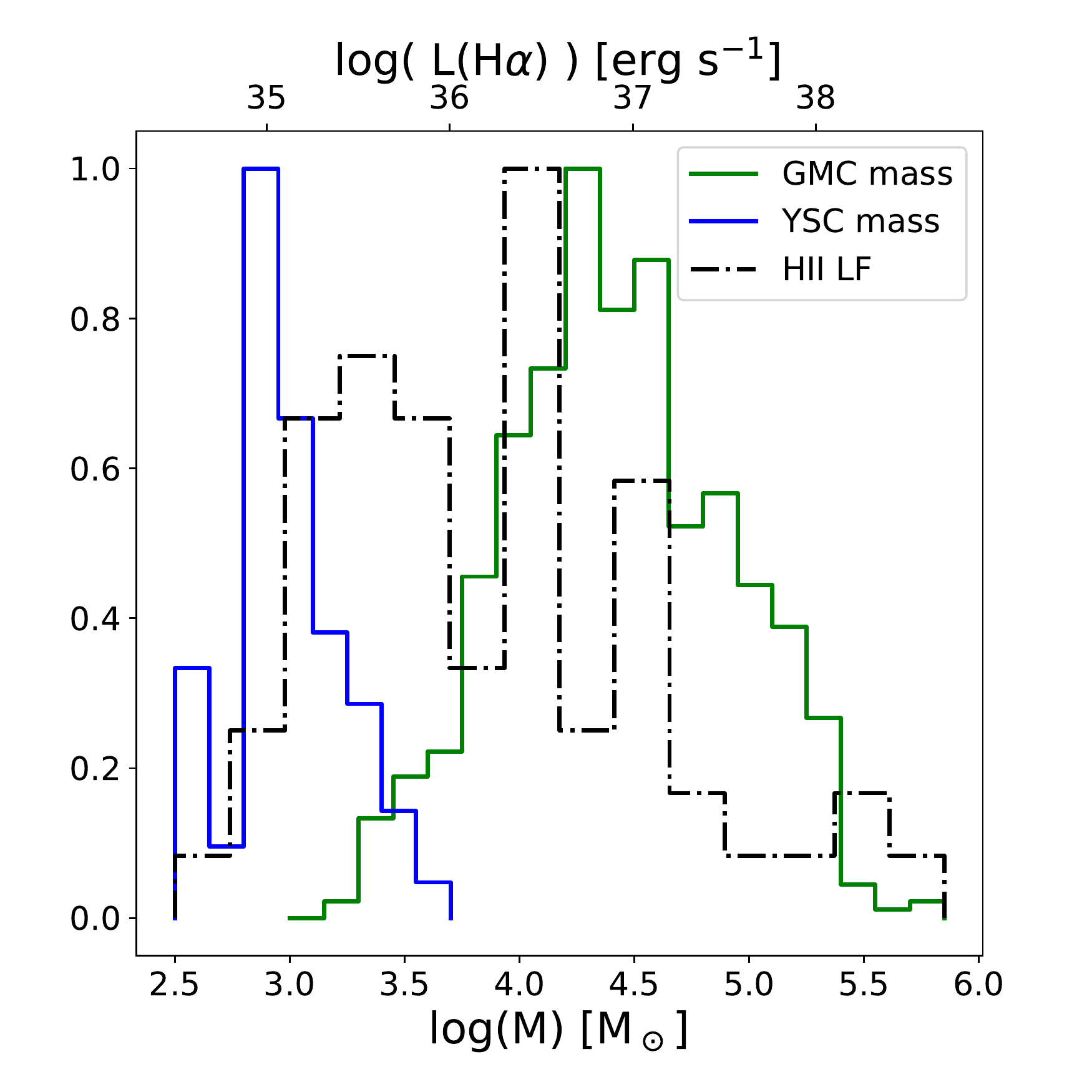}
\caption{Distribution of YSC mass (clusters with $t < 10$ Myr, in blue), GMC mass (in green) and \ion{H}{ii} regions luminosity (in black, reddening corrected) in our FoV. YSC masses are best estimates computed from the median of the corresponding PDF (see Sect.~\ref{section:ysc_properties}). The histograms are normalised to a peak value of 1.}
\label{fig:mass_distr_gmc_ysc}
\end{figure}

\subsection{Stellar census}
We present here the full census of the stellar population in our FoV. Figure~\ref{fig:stellar_census} shows a map of the H$\alpha$ emission line obtained from the integration of the MUSE stellar continuum subtracted datacube in the rest-frame wavelength range $6559 - 6568$~\AA{}. Overlaid on the H$\alpha$ emission, we show the position of candidate O stars from Lee et al. (in prep., white circles), YSCs from LEGUS (green triangles) and of Wolf Rayet (WR) stars that we spectroscopically identified in Paper~\textsc{i} (listed in Table~\ref{table:wr_stars}; blue crosses).
We also show the contours of \ion{H}{ii} regions identified in Paper~\textsc{i} (in black, corresponding to a flux brightness cut in H$\alpha$ = $\SI{6.7e-18}{\erg \per \second \per \centi \metre \squared}$spaxel$^{-1}$) and of GMCs (light blue filled contours). 
The total stellar census in our FoV consists of 651 O stars, 93 YSC, and 9 WR stars.

From Fig.~\ref{fig:stellar_census}, we observe that both YSCs and candidate O stars are abundant over the entire FoV. However, we note that several of the smaller and medium-bright \ion{H}{ii} regions do not seem to host any cluster. Overall, we find
$\sim$ 49\% field\footnote{Throughout the paper, we refer to stars and clusters located outside \ion{H}{ii} regions as `field' objects.} clusters (35\% if considering only clusters with median age $<$ 10 Myr, as motivated in Sect.~\ref{section:fesc_individual}) and
$\sim$ 38\% field candidate O stars. 
We note that the numbers of O-stars quoted here are not corrected for completeness, so that the latter fraction could in part be due to the high contamination rate of catalogue (see Sect.~\ref{section:data}).
We observe that WR stars are populating the \ion{H}{ii} regions, with the exception of one object (located in the upper left corner and catalogued as WR 3 in Table~\ref{table:wr_stars}). \citet{rate20} recently found that isolated WR stars are not an exception, and amount to at least 64\% in the Milky Way \citep[as observed in][]{gaia_DR2}. The authors explore the possible origin of field WR stars in simulations, finding that the most frequent mechanisms are the formation in a low density association that then expands during the star lifetime and the ejection from a cluster through dynamical interaction or binary disruption. This has been observed in galactic star clusters \citep[e.g. Westerlund 2 and NGC 3603][]{drew18, drew19}.
The fact that WR 3 is located $\sim$ 100 pc away from a nearby HII-region cluster (see Fig.~\ref{fig:stellar_census} and~\ref{fig:sii_oiii}) could support the latter formation scenario.
In order to investigate this hypothesis, we have computed the minimum velocity needed for the star to be ejected from the nearby cluster. We assume an age range of (4 $\pm$ 1) Myr for the WR star\footnote{We note that, despite the spectral information available, the age also depends on the mass of the progenitor O stars and on rotation effects.}.
Assuming that the star has been ejected immediately after birth, we find $v_{proj, min} = 25_{-8}^{+5}$ km/s, which meets the standard runaway criterium of $v > 25$ km/s \citep[e.g.][]{drew18, portegieszwart00}.

\subsection{YSCs age, mass, and ionising flux distribution}
\label{section:ysc_properties}
In this subsection, we investigate the physical properties of the YSCs, namely their age, mass, and ionising photon flux. The latter, $Q(H^0)$ [s$^{-1}$], is a measure of the number of Lyman continuum (LyC) hydrogen-ionising photons ($h\nu > 13.6$ eV) emitted per unit time.

For this purpose, we use the stochastic stellar population synthesis code \textsc{slug} \citep[][v2]{dasilva12,krumholz15a}. 
\textsc{Slug} implements a Monte Carlo technique by randomly drawing stars from an input initial mass function (IMF). This approach allows one to account for the effect of stochastic sampling of the IMF, which is especially important when populating low mass clusters \citep[see e.g.][]{hannon19}.
This is of a particular relevance in our target: NGC 7793 has a very low SFR $\sim$ 0.1 $M_\sun$ yr$^{-1}$ \citep{calzetti15}, resulting in \ion{H}{ii} regions hosting low mass ($\sim$ a few 1000s $M_\sun$) clusters.
The \textsc{SLUG} software package includes the \texttt{cluster\_slug} tool~\citep{krumholz15a}, that implements Bayesian inference techniques to calculate posterior probability distribution functions (PDFs) of physical parameters of star clusters based on  observed cluster photometry, as described in \citet{krumholz15b}.
This is achieved by comparing the data to a large library of simulated clusters. We modified \texttt{cluster\_slug} to also return the ionising photon luminosity $Q(H^0)$ as a further output parameter and ran the code on HST broadband photometry of the YSCs. We consider age $t$, mass $M$ and visual extinction $A_V$ as free parameters, and assume a flat prior in $A_V$ and in $\log t$, and a $\log(M) \sim 1/M$ prior on the mass. We use the library of mock star clusters described in~\citet{ashworth18}, a Milky Way extinction law by \citet{fitzpatrick99} and the non-rotating solar metallicity stellar population models from \citet{genevamodels12}. The abundance map derived in Sect.~\ref{section:metallicity} confirms that NGC 7793 has a metallicity close to solar \citep[see also][]{pilyugin14}; the same metallicity was also assumed in the \textsc{slug} models of \citet{krumholz15b}. We discuss the impact of including binary stars and the effect of stellar rotation on the resulting physical properties in Sect.~\ref{section:discussion}.

\par Figure~\ref{fig:ysc_distr_FOV} shows the spatial distribution across the FoV of YSC ages (left panel) and masses (right panel). We are showing here as best estimate the median of each cluster PDF.
As remarked in \citet{krumholz15b}, reducing a full PDF to a single point inevitably leads to some imprecision, especially if the PDF is multi-peaked. We follow the method tested in \citet{krumholz15b} and we refer the interested reader to this work for evaluations of the methodology and effects on the recovered cluster physical parameters.
In Fig.~\ref{fig:ysc_agemass_hist}, we show the age, mass, and ionising flux PDF for YSCs located inside \ion{H}{ii} regions (in blue) and for field clusters (in orange). Thin and thick lines in the plot indicate, respectively, the PDFs of single clusters and the combined PDFs obtained by summing the fractional probability of each cluster in every age, mass, and flux bin, and re-normalising the resulting distribution.
% Age
In the case of the age, both the single and combined PDFs are (most often) multi-peaked; the main peak of the combined PDF
is located at $t \sim 5$~Myr, both for field and \ion{H}{ii} region clusters. However, we see that the PDF of field clusters flattens towards older ages, while the probability distribution of \ion{H}{ii} region clusters is more peaked at ages below 10 Myr, that is clusters located in \ion{H}{ii} regions have a higher probability to be young and potentially ionising sources.
% Mass
The mass distribution is single-peaked for both categories of clusters, with a tail towards higher masses. The peak of the distribution is located at $\log(M) = 3.0$ and 3.4~$M_\odot$ for \ion{H}{ii} region clusters and field clusters respectively.
% Q(H0)
The $Q(H^0)$ distribution is, as the age, multi-peaked (\text{both the single and combined PDFs}) with two main modes (both for \ion{H}{ii} region and field clusters) at $\log(Q(H^0)) = 44.1$ and 48.8 s$^{-1}$.
Nevertheless, we notice that clusters in \ion{H}{ii} regions have a higher probability to be associated with the high $Q(H^0)$ value, while the opposite is true for field clusters. In general however, the inferred masses and $Q(H^0)$ are on the low range for YSCs: for a reference, the peak at $\log(Q(H^0)) = 48.8$ s$^{-1}$ corresponds to the emission of a single O7.5 star \citep{martins05}. We comment more on the mass distribution below (Fig.~\ref{fig:mass_distr_gmc_ysc} and associated text).

% Overall
In synthesis, we observe that YSCs in \ion{H}{ii} regions have a higher probability to be younger, slightly lower mass, but a more significant source of ionising photons than clusters in the field. 
These trend can easily be visualised in the plots of Fig.~\ref{fig:ysc_distr_FOV}, where we colour code the positions of the clusters in the FoV according to their median ages and masses.

\par The age range observed for the clusters within the \ion{H}{ii} regions is in agreement with studies of YSCs in local galaxies \citep{whitmore11, hollyhead15, hannon19, grasha19}, that find that clusters emerge from their natal gas in $< 4 - 5$ Myr, but that the process can start as early as $2 - 3$ Myr. Also simulations of cluster formation including solely the effect of photoionisation and stellar winds \citep{dale14, bending20}, have shown that within 3 Myr, and before the onset of SN, several clusters have vacated some channels with low gas density. These channels are easily ionised; therefore ionising radiation can escape from the \ion{H}{ii} regions even if star formation is still taking place. In Fig.~\ref{fig:stellar_census} we observe that a significant amount of molecular gas is still detectable in the \ion{H}{ii} regions hosting YSCs, in agreement with the findings of \citet{dale14}.

\par Finally, in Fig.~\ref{fig:mass_distr_gmc_ysc} we show the total demographics of GMCs (in green), YSCs with age $< 10$ Myr (in blue) and \ion{H}{ii} regions (in black) within our FoV, highlighting the various phases of the star formation cycle. We observe that GMCs, hosting the molecular gas that will give rise to the next generation of stars, span a rather low range of masses compared to what observed in other local galaxies \citep[$\log M_{GMC} \sim 4.5 - 8 M_\odot$,][]{hughes16}.

The collapse of the clouds results in turn in low-mass YSCs. While there is not a one-to-one correlation between the mass spectrum of GMCs and YSCs, we observe that overall the two distributions would suggest an integrated SFE of a few percent of what is typically observed in local galaxies \citep[see e.g. the review by][]{krumholz19}.
The bulk of the YSCs have masses around 1000 $M_\odot$, and the distribution reaches up to $\sim 10^{3.5}~M_\odot$. These mass distributions are typical of galaxies with low SFR, such as M31 \citep{larsen09, johnson17, adamo20}.
If we consider that the ionising feedback from a cluster of 1000 $M_\odot$ gives rise to an H$\alpha$ luminosity
of $\log(L) \sim 37.3$ erg s$^{-1}$,\footnote{\textsc{Starburst99} \citep{leitherer99} estimate assuming a \citet{kroupa01} IMF in the stellar mass range $0.1-120~M_\odot$.}
then the observed luminosity distribution is consistent with the mass distributions of the YSCs in the FoV. If compared to studies of \ion{H}{ii} regions luminosity functions in the local universe, our sample occupies the low-L end of the luminosity function of local spirals \citep[$\log(L ) \sim 37 - 40$ erg s$^{-1}$,][]{kennicutt89a}. Overall, the low masses and luminosity observed are in agreement with the flocculent morphology and low SFR of our target.

\begin{figure*}%[ht]
%\center
\includegraphics[width=8.3cm]{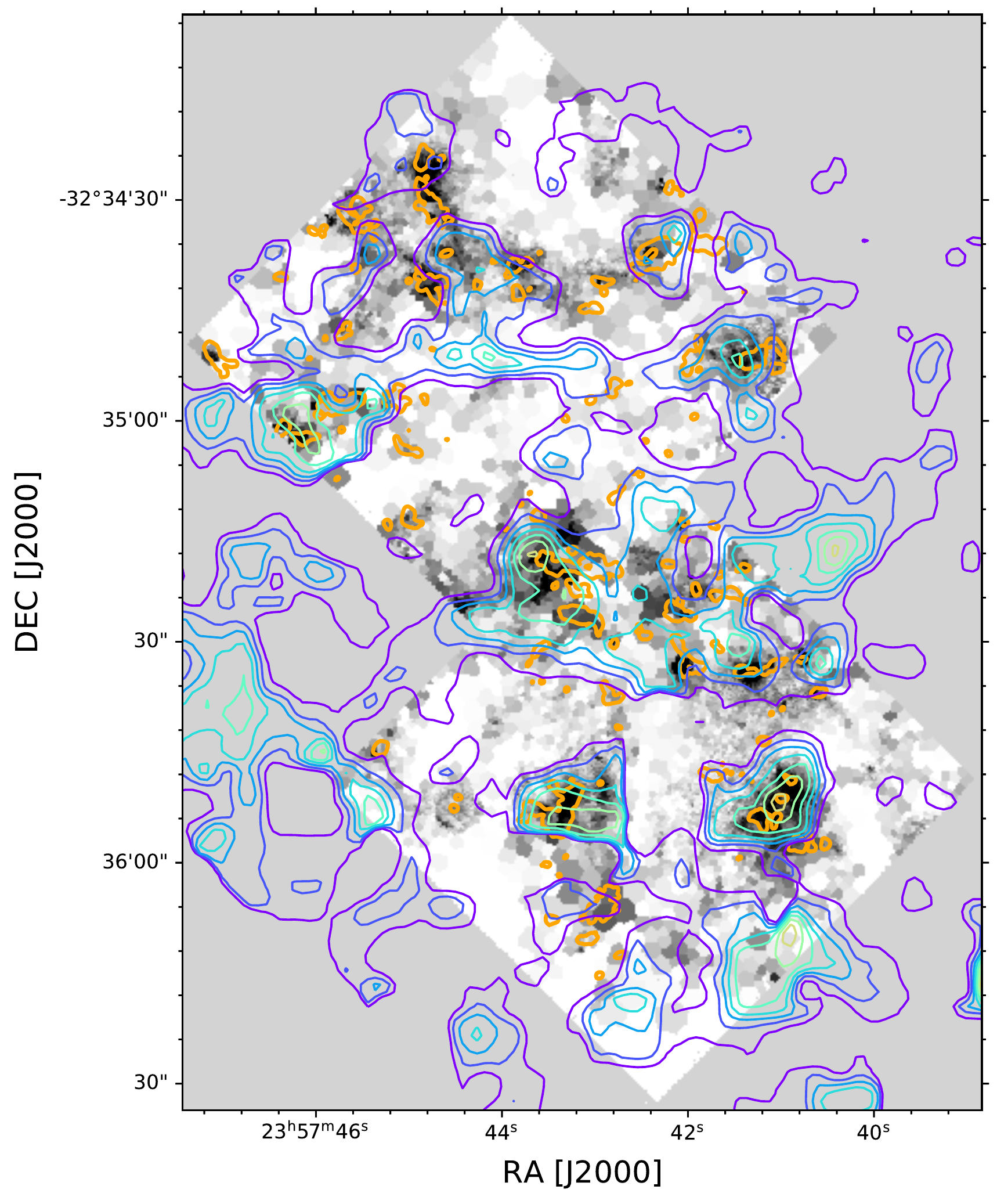}
\includegraphics[width=9.6cm]{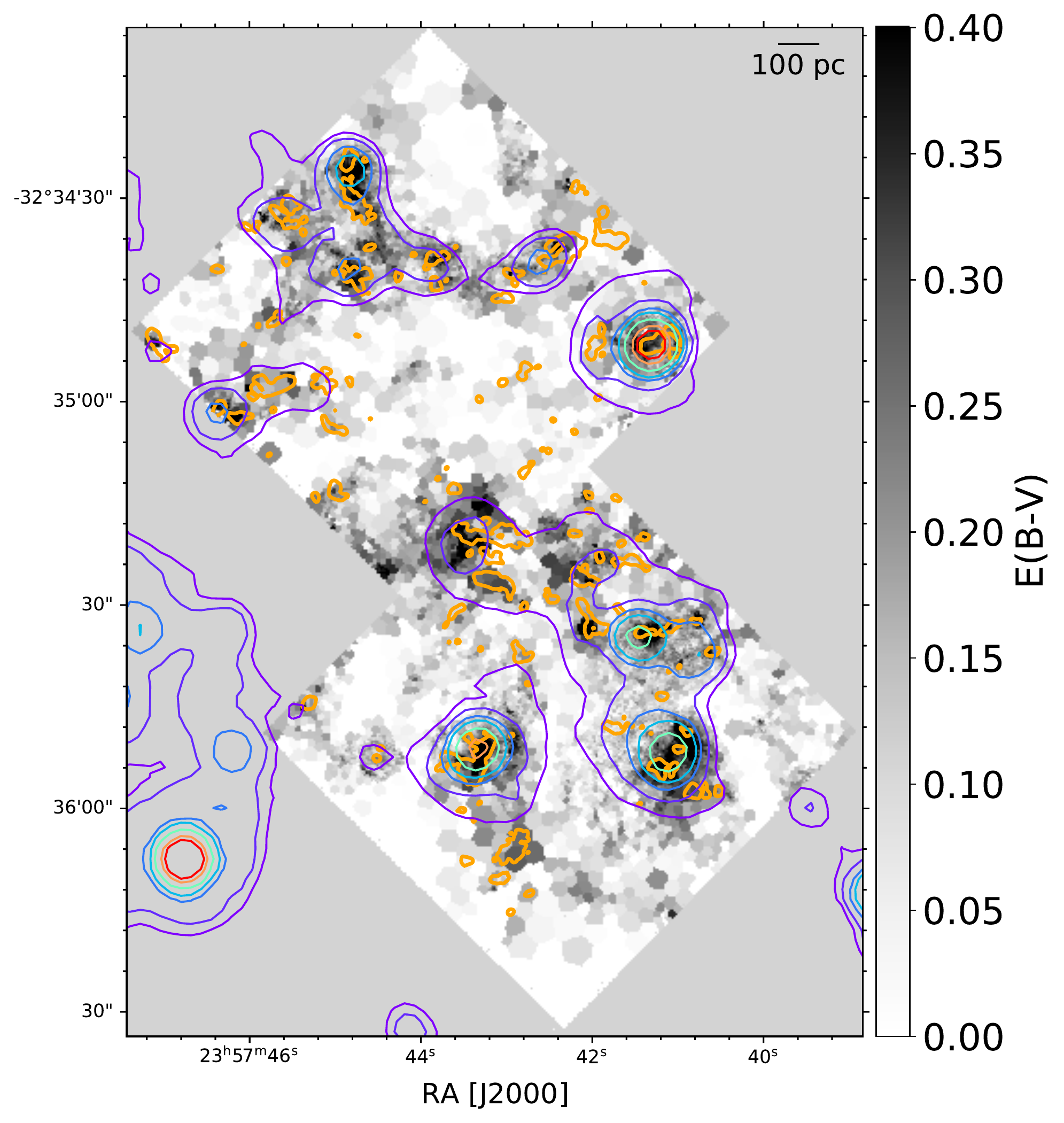}
\caption{Extinction map derived from the ionised gas observed with MUSE assuming an intrinsic H$\alpha$/H$\beta$ ratio (from Paper~\textsc{i}) compared to: \textit{Left panel}: contours of the extinction map derived by~\citet{kahre18} from the individual stellar extinctions obtained from HST data. The contours correspond to values of $E(B-V) = 0.2 - 0.5$, in steps of 0.033. \textit{Right panel}: 24~$\mu$m emission from Spitzer/MIPS. The contours correspond to a flux of [1.5, 2, 3, 4, 6, 9, 11] MJy/sr. The thick orange contours indicate the position of GMCs.}
\label{fig:extinction_map}
\end{figure*}

\section{Comparison between stellar and ionised gas extinction with maps of cold dense gas and hot dust emission}
\label{section:extinction}
The HST, MUSE, ALMA, and Spitzer/MIPS\footnote{Publicly available in the NED archive \url{http://ned.ipac.caltech.edu}.} coverage gives us the rare opportunity of comparing extinctions as traced by stellar reddening and the ionised gas phase to the distribution of the cold dense gas and the hot dust emission.
In Paper~\textsc{i}, we have derived an extinction map from the MUSE H$\alpha$/H$\beta$ ratio using \textsc{pyneb} \citep{luridiana15}. Hereby, we assumed a theoretical H$\alpha$/H$\beta_{int}$ = 2.863 and we obtained the observed ratio by binning the stellar continuum subtracted cube with the Voronoi technique of \citet{cappellari03} to a S/N = 20 in H$\beta$ and fitting both lines with a single Gaussian profile.
Here, we compare the gas extinction map derived in Paper~\textsc{i} with the map derived by~\citet{kahre18} from the individual stellar extinctions obtained from HST, and correlate the two maps with the position of GMCs traced by ALMA.
Figure~\ref{fig:extinction_map} (left panel) shows overlaid on the gas extinction map the contours of the weighted average map with adaptive resolution from~\citet{kahre18}. The latter was derived by spatially binning the HST data in the smallest possible regions (of varying size between 1 and 10 arcsec$^2$) hosting at least 10 O stars. In order to compare the two maps, we have converted the colour excess E(V$-$I) to E(B$-$V) assuming a Milky Way extinction law with $R_V = 3.1$, as in \citet{kahre18}.
We observe a clear correspondence between the two maps, despite the difference in binning and in the method used to derive the extinction. In general, regions in which we measure a high extinction also feature a comparable stellar extinction, although the position of the extinction peaks is slightly shifted in the two maps. Conversely, the ionised gas extinction seems to miss some of the regions of enhanced stellar extinction. We notice that these regions lie outside the bright ionised gas, where H$\beta$ is poorly detected and the ionised gas might not be a sensitive enough tracer of extinction.
However, we observe that the gas extinction seems to better correlate with the position of the GMCs (orange contours in Fig.~\ref{fig:extinction_map}). We also note that there seems to be a better agreement between gas extinction peaks and GMC position within \ion{H}{ii} regions.
\par
In the right panel of Fig.~\ref{fig:extinction_map}, we furthermore show overlaid on the gas extinction map contours of 24~$\mu$m emission from Spitzer/MIPS. The 24~$\mu$m emission is tracing hot dust in the galaxy, and inside the \ion{H}{ii} regions it is indicative of where star formation is taking place. We observe a good correlation between gas extinction, GMC position and 24~$\mu$m emission, indicative of the fact that, in HII regions, the dust is well mixed with gas. We comment more on this fact in relation with the effect of removal of LyC photons by dust prior to their absorption by neutral hydrogen in Sect.~\ref{section:discussion}.

\begin{figure*}%[ht]
%\center
\includegraphics[width=9.2cm]{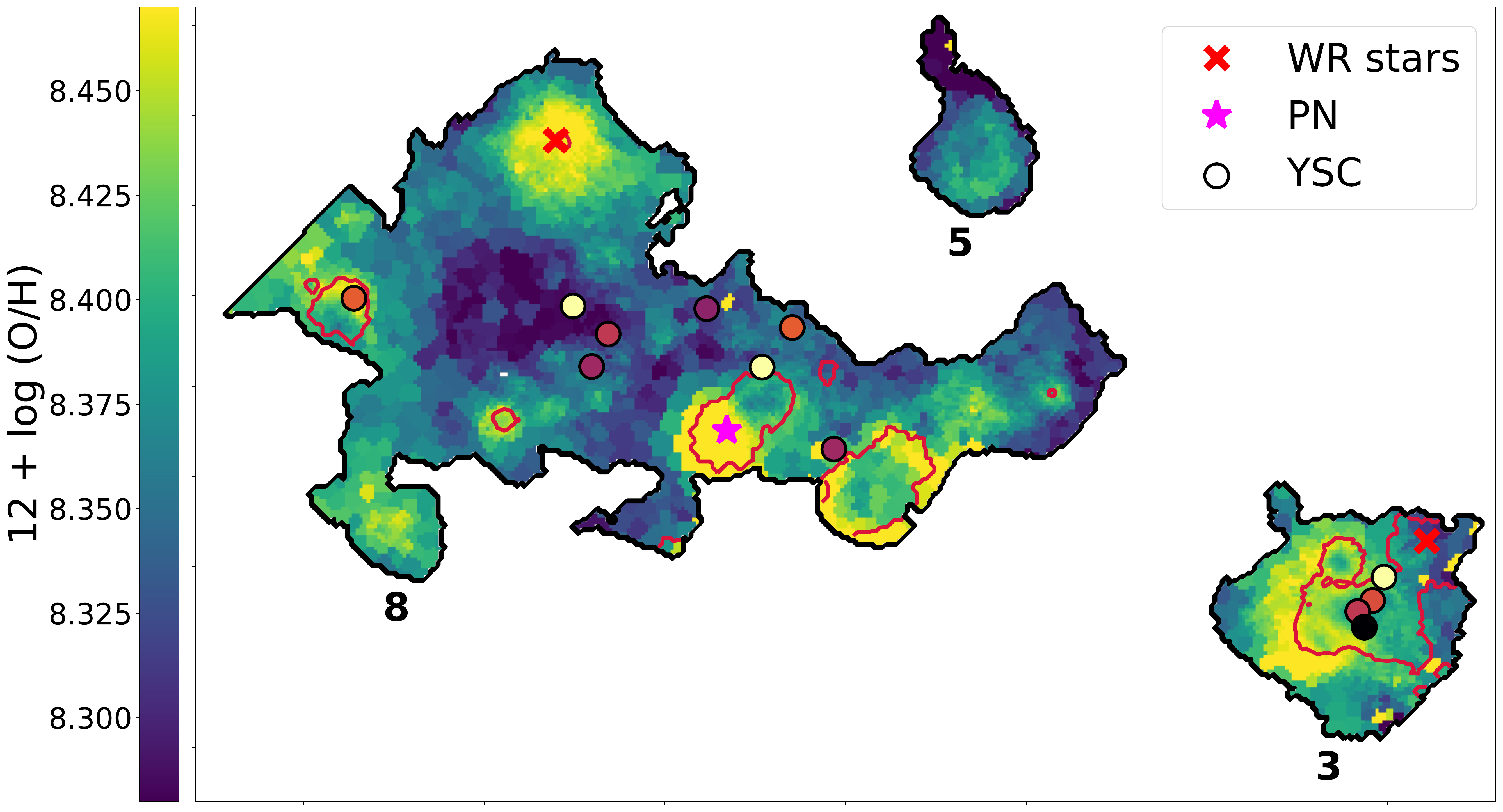}
\includegraphics[width=9.0cm]{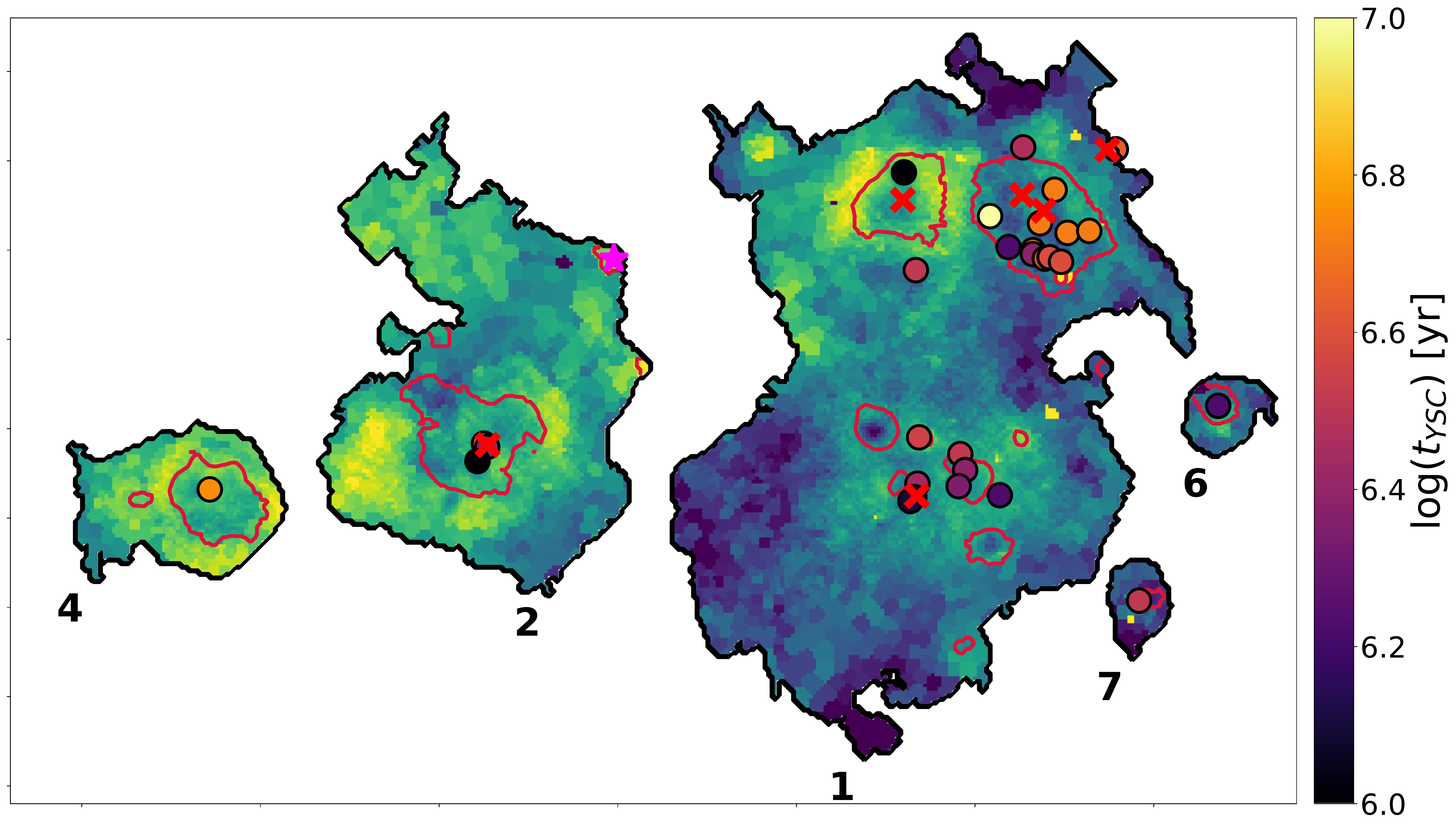}
\caption{Oxygen abundance for the subset of \ion{H}{ii} regions labelled in Fig.~\ref{fig:sii_oiii}, determined with the $S$-strong line calibration of \citet{pilyugin16}. The location of YSCs is indicated with filled circles, colour-coded based on their age (best estimates derived from the median of each PDF). Red crosses and purple stars correspond to the position of WR stars and PNe, respectively. Dark red contours indicate a ratio of [\ion{S}{ii}]/[\ion{O}{iii}] = 0.5 (see Fig.~\ref{fig:sii_oiii}).}
\label{fig:logO_H_map}
\end{figure*}

\section{Mapping the Oxygen abundance within the \ion{H}{ii} regions}
\label{section:metallicity}
\begin{figure*}%[ht]
\center
\includegraphics[width=14cm]{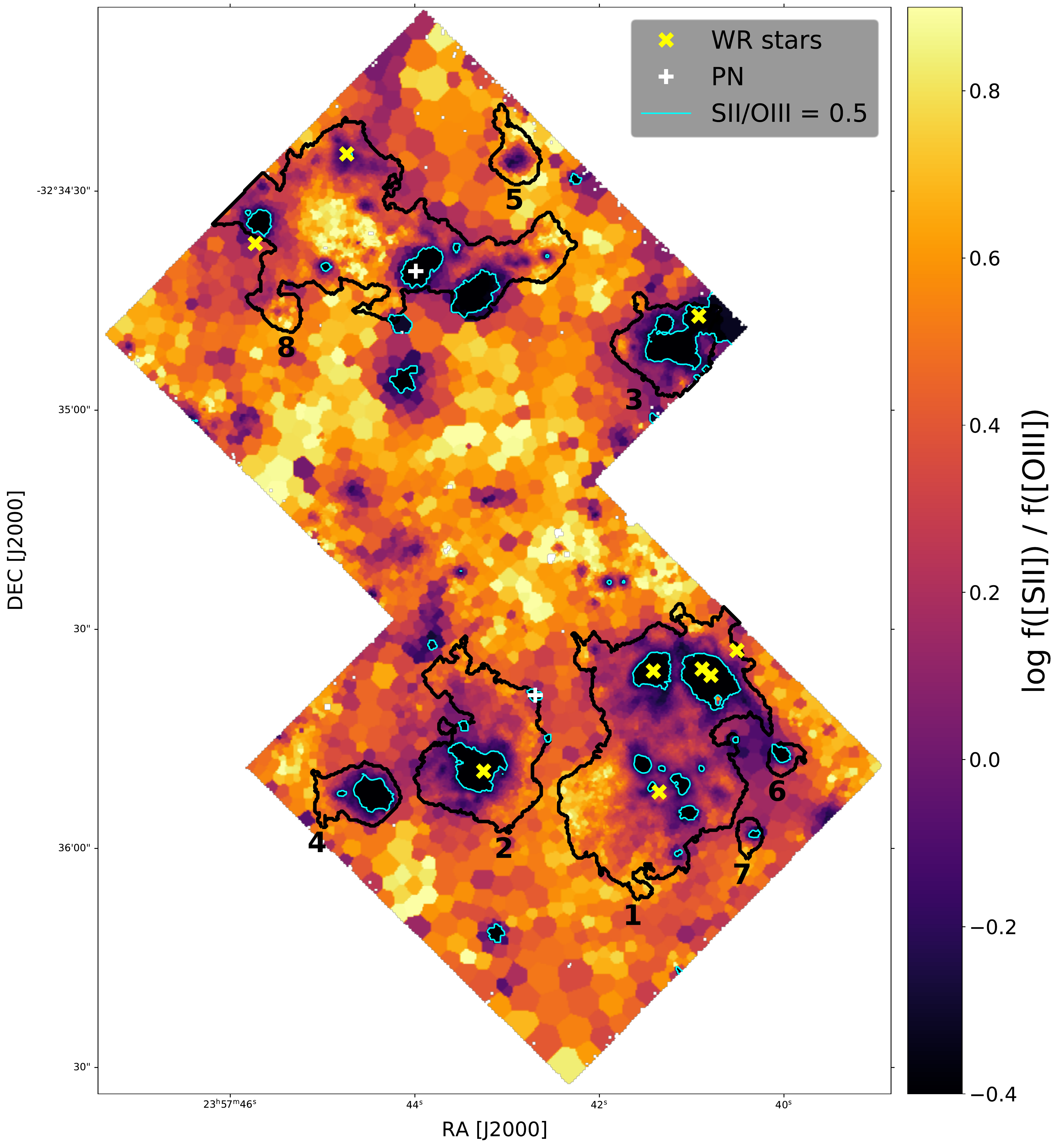}
\caption{[\ion{S}{ii}]6716,31/[\ion{O}{iii}]4959,5007 ratio map (reddening corrected), tracing the ionisation structure of the gas.
The cyan contours indicate a ratio of 0.5, which we use as fiducial limit for the transition to optically thick gas.
Numbered black contours correspond to the \ion{H}{ii} regions analysed in Sect~\ref{section:budget}. The position of WR stars (yellow crosses) and planetary nebulae (white plus symbols) is also indicated.}
\label{fig:sii_oiii}
\end{figure*}

We estimate an oxygen abundance in the \ion{H}{ii} regions using the strong-line method of \citet{pilyugin16}.
In Paper~\textsc{i}, we derived electron temperature and density from the [\ion{S}{iii}]6312/9069 and [\ion{S}{ii}]6716/6731 line ratios\footnote{As remarked in Paper~\textsc{i}, the [NII]5755 `auroral' line used for temperature diagnostics falls  in the wavelength range blocked out in the MUSE/AO mode.}; however, due to the weakness of the [\ion{S}{iii}]6312 line and to the fact that the [\ion{S}{ii}] ratio is largely insensitive to $n_e < 30$ cm$^{-3}$, we were only able to obtain a coarse $T_e$ map and marginally constrained $n_e$, preventing us from measuring direct abundances.
The \citet{pilyugin16} method bases on ratios of strong lines calibrated on a sample of $\sim$ 300 \ion{H}{ii} regions. Here we make use of the $S$ calibration, based on the following three line ratios:
$$N_2 = (I_{[\ion{N}{ii}]}\lambda6548 + I_{[\ion{N}{ii}]}\lambda6584) / I_{H\beta},$$
$$S_2 = (I_{[\ion{S}{ii}]}\lambda6716 + I_{[\ion{S}{ii}]}\lambda6731) / I_{H\beta},$$
$$R_3 = (I_{[\ion{O}{iii}]}\lambda4959 + I_{[\ion{O}{iii}]}\lambda5007) / I_{H\beta}.$$
We use the upper branch of the calibration, valid for $\log N_2 \gtrsim -0.6$, as our data lie well above this limit throughout the FoV. The line ratios are determined from the reddening corrected fluxes, which have been obtained by binning the data with the Voronoi technique to a S/N = 20 in H$\beta$ as described in Sect.~\ref{section:extinction}.
Figure~\ref{fig:logO_H_map} shows the resulting metallicity variation in a subset of the \ion{H}{ii} region sample, for which we construct a detailed photoionisation budget in Sect.~\ref{section:budget}. The exact location of these regions in the FoV is shown in Fig.~\ref{fig:sii_oiii}. We observe that the abundance ranges from $12 + \log(O/H) \sim 8.25$ to 8.50, with a median value of 8.37.
This range of values is in agreement with the strong-line estimate of \citet{pilyugin14} for NGC 7793 and with the metallicity range inferred in Paper~\textsc{i} when comparing our data with the models of star forming galaxies from \citet{levesque10} in `BPT' diagrams \citep{baldwin81}.

\par In small and isolated \ion{H}{ii} regions, such as regions 4, 5, 6, and 7, we notice that the oxygen abundance varies of about 0.1 dex at most. The largest \ion{H}{ii} region complexes, such as regions 1, 2, 3, and 8 show larger gradients and pockets of enriched gas coincident or arranged in shell-like structures around clusters and massive stars\footnote{We omit to plot the position of O stars, which are also a source of enrichment, to avoid overcrowding and facilitate the visualisation of the abundance variation. We refer the reader to Fig.~\ref{fig:stellar_census}.}.
This is particularly visible in region 4, where the YSC is surrounded by an enriched shell, and in regions 6 and 8 (centre left and bottom right) where we estimate a local enhancement in the oxygen abundance at the location of YSCs.
We furthermore observe that some of the WR stars are associated with enriched shells or pocket of ionised gas: this is the case in region 1 (WR2) and 8 (WR4), likely due to nitrogen-enriched gas expelled by their stellar winds. In other cases, we do not observe any enrichment surrounding the stars: this is especially visible in region 3 and in the north-west, and south of region 1.
Finally, we observe a large enhancement of the abundance at the location of one of the planetary nebulae identified in Paper~\textsc{i} (PN2).

\par We would like to point out that, although such indirect methods are generally calibrated for unresolved \ion{H}{ii} regions, they have recently been applied also to study abundance variations within resolved \ion{H}{ii} regions \citep[e.g.][]{james16, mcleod19}.
Using narrow-band HST maps, \citet{james16} show chemical inhomogeneity ($\sim$ 0.1 dex) on scales smaller than 50 pc within the star forming region Mrk 71. \citet{mcleod19} also report abundance variations and structures similar to what we observe here in two star-forming regions of the Large Magellanic Cloud, resolved at scales of a few parsec. 
However, as remarked, for example, by \citet{mcleod19}, abundances derived with a strong line method can have a secondary non-trivial dependence on temperature and density, resulting for instance in falsely low abundances in regions with a high ionisation state~\citep{ercolano12,mcleod19,mcleod16a}. In Fig. \ref{fig:logO_H_map}, we show contours of [\ion{S}{ii}]/[\ion{O}{iii}] = 0.5 (dark red), delimiting strongly ionised regions in the nebulae (see Sect.~\ref{section:optical_depth}). We observe indeed a close correspondence between the contours and areas of low metallicity in all the regions, pointing to the fact that a degeneracy between oxygen abundance and ionisation conditions is likely playing a role closest to the young massive stars. 
In the future, IFU having a spectral and spatial capability comparable to MUSE, but covering a bluer wavelength range \citep[e.g. BlueMUSE,][]{bluemuse}
will help us to better understand abundances in resolved \ion{H}{ii} regions by addressing all such degeneracies.

\begin{figure}%[ht]
\center
\includegraphics[width=8cm]{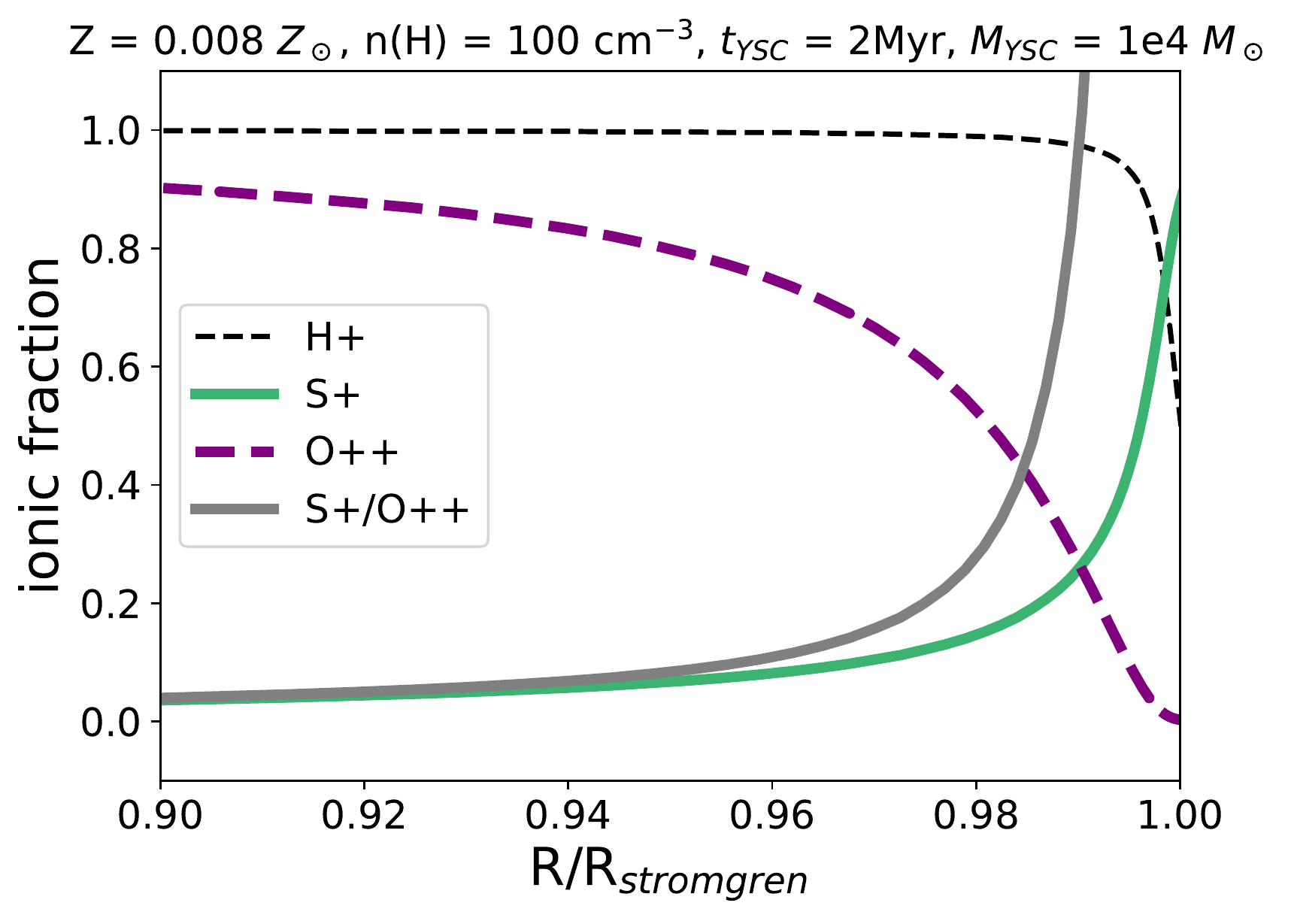}
\caption{\textsc{cloudy} model illustrating the ionisation structure of an ideal ionisation-bound nebula surrounding a 2 Myr YSC of $M = 10^4~M_\odot$ at the metallicity of NGC 7793. The radial variation in ionic fraction of \ion{H}{i} (dashed black), \ion{O}{iii} (dashed purple), \ion{S}{ii} (solid green) and the \ion{S}{ii}/\ion{O}{iii} ratio (solid grey) are shown. The nebula consists of two main zones: an inner zone dominated by \ion{S}{iii} and \ion{O}{iii}, and an outer zone dominated by \ion{S}{ii} and \ion{O}{ii}.}
\label{fig:cloudy_model}
\end{figure}

\begin{figure}%[ht]
\center
\includegraphics[width=7.8cm]{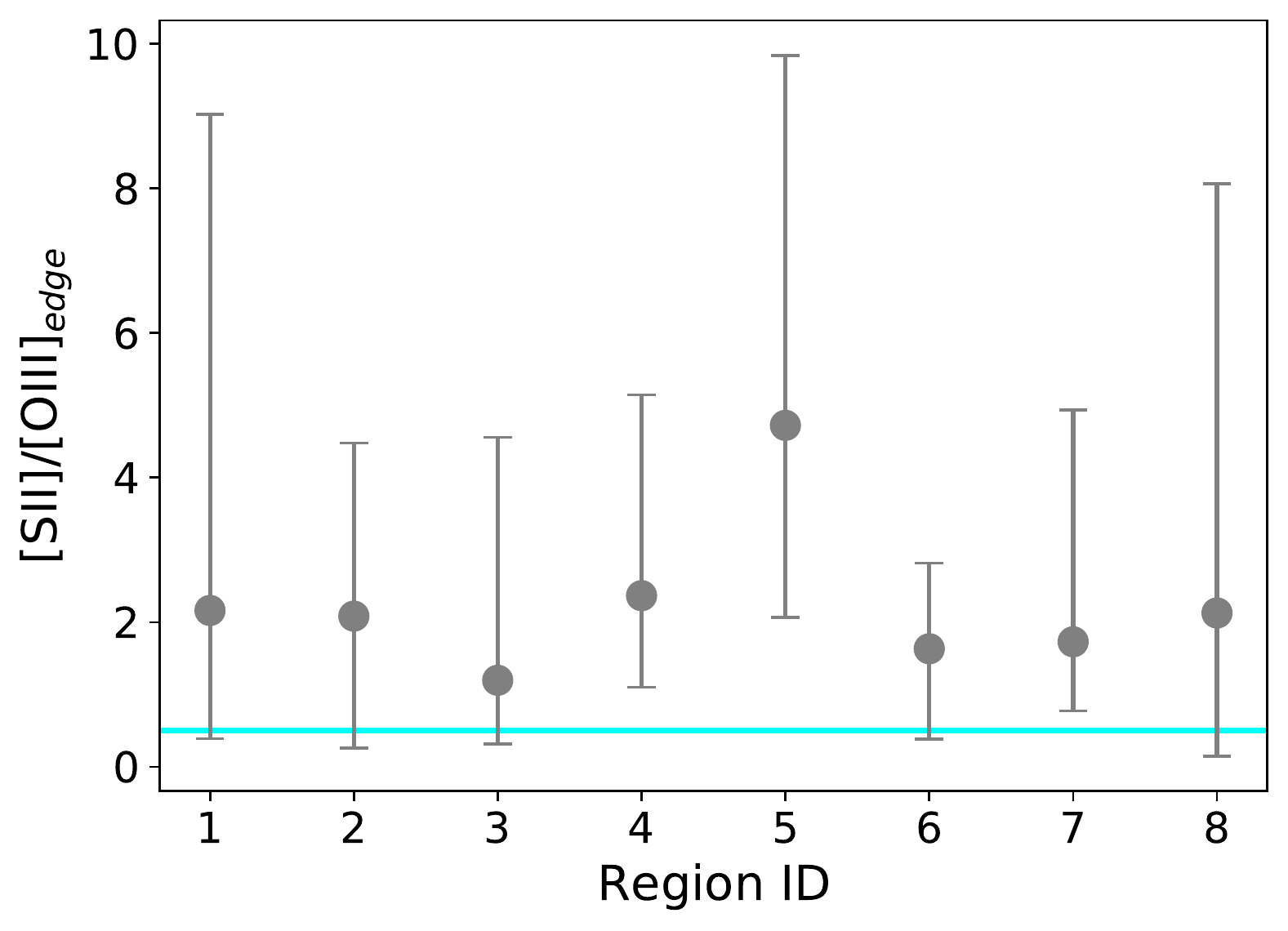}
\caption{Values of [\ion{S}{ii}]6716,31/[\ion{O}{iii}]4959,5007 along the contour of the \ion{H}{ii} regions labelled in Fig.~\ref{fig:sii_oiii}. The solid points and error bars show, respectively, the median and the range in values spanned by the ratio. The cyan line indicates a ratio of 0.5, which we use as fiducial limit for the transition to optically thick gas. Regions for which the range in [\ion{S}{ii}]/[\ion{O}{iii}] extends below this limit are classified as featuring optically thin channels (CH); all other regions are classified as ionisation bounded (IB).}
\label{fig:hii_class}
\end{figure}

\begin{figure}%[ht]
\center
\includegraphics[width=9cm]{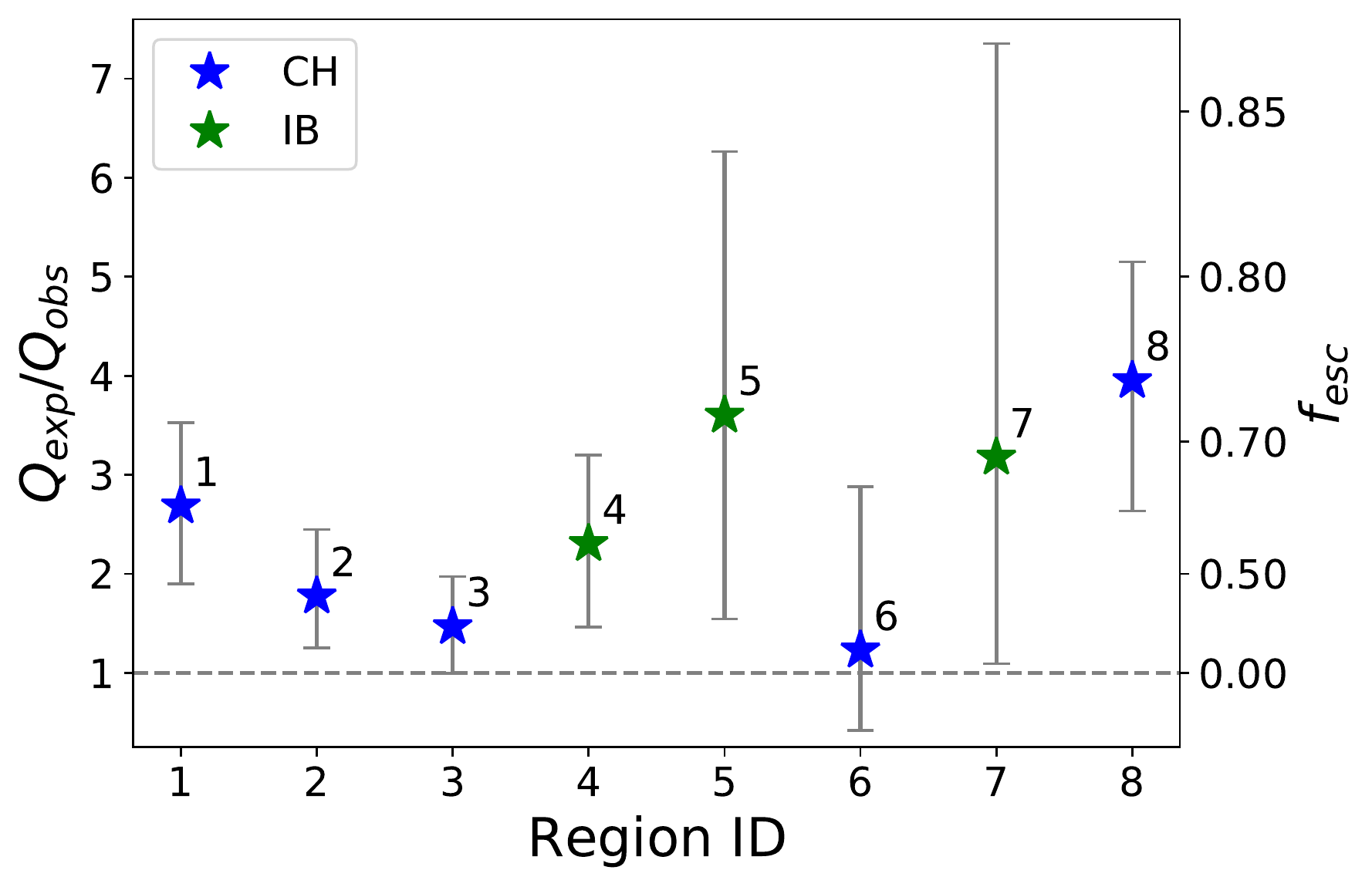}
\caption{Ratio of expected to observed ionising luminosity $Q(H^0)$ for the regions labelled in Fig.~\ref{fig:sii_oiii} versus their visual appearance: ionisation bounded regions (IB, green stars) vs regions featuring optically thin channels (CH,  blue stars). The dashed horizontal line indicates an $f_{esc} = 0$.}
\label{fig:fesc_vs_appearance}
\end{figure}

\section{Ionisation structure of the \ion{H}{ii} regions}
\label{section:optical_depth}

The ionisation structure of a nebula can be studied through the relative emission of two ions with different ionisation potential. The structure of an ideal ionisation bounded \ion{H}{ii} region is shown in Fig.~\ref{fig:cloudy_model}.
The figure illustrates a model generated with \textsc{cloudy} for a nebula at the metallicity of our target ($Z \sim 0.008$, see Sect.~\ref{section:metallicity}), surrounding a 2 Myr YSC with a mass of $10^4~M_\odot$. The simulation is run until the gas interface reaches an optical depth $\tau = 2$.
Two zones can be distinguished in the nebula: a central zone with a higher ionisation state, dominated by S++ and O++ and an external zone with a lower ionisation state, where these ions are singly ionised. The ratio of S+/O++ (thick grey) can then be used as a proxy for the optical depth of the nebula, that progresses from optically thin in the inner zone to optically thick in the outer zone, with a sharp cutoff at the Strömgren radius.
In reality, the inner zone is often not, or only partially embedded in the optically thick outer envelope, so that a fraction of hydrogen-ionising photons are not absorbed and can escape into the ISM.
Studying the spatial extent of the two zones through emission from their most abundant ions can provide us with information about the ionisation structure of an \ion{H}{ii} region. \citet{pellegrini12} exploited this dependency to develop the ionisation parameter mapping (IPM) method and applied it to study the optical depth of \ion{H}{ii} regions in the Large and Small Magellanic Clouds (LMC and SMC) with narrow-band photometry in [\ion{S}{ii}], [\ion{O}{iii}] and H$\alpha$. Regions were classified based on the fraction of the central high-ionisation region being surrounded by a low-ionisation envelope in an [\ion{S}{ii}]/[\ion{O}{iii}] emission map. Regions completely embedded in the envelope are labelled as optically thick, regions with a low [\ion{S}{ii}]/[\ion{O}{iii}] throughout as optically thin and those with an in between morphology as `blister nebulae'. 
Each morphological class was then assigned an escape fraction; we discuss the results of this approach in the light of our work in Sect.~\ref{section:discussion}.

\par Figure~\ref{fig:sii_oiii} shows a map of the [\ion{S}{ii}]6716,31/[\ion{O}{iii}]4959,5007 ratio in our FoV. The map has been reddening corrected assuming an intrinsic H$\alpha$/H$\beta$ ratio of 2.863, as described in Paper~\textsc{i}. The data have been Voronoi binned to a S/N = 20 in H$\beta$. The numbered black contours in Fig.~\ref{fig:sii_oiii} indicate the subset of the \ion{H}{ii} regions - out of the full sample identified in Paper~\textsc{i} and shown in Fig.~\ref{fig:stellar_census} - that we study in detail in Sect.~\ref{section:budget}.
The bar shows increasing ratios of [\ion{S}{ii}]/[\ion{O}{iii}], from black to yellow, corresponding to an increasing optical depth of the medium (or, a decrease in its transparency to ionising photons). A ratio of 1 (purple) indicates where the abundance of S+ prevails over O++, as shown in the model in Fig.~\ref{fig:cloudy_model}\footnote{We caution however that Fig.~\ref{fig:cloudy_model} shows the ratio of the ionic fractions, which does not compare directly to a line emissivity ratio.}.

% DIG and Morphological classification of regions
\par We observe that the DIG has a lower ionisation state than the \ion{H}{ii} regions (higher ratios of [\ion{S}{ii}]/[\ion{O}{iii}]),
as we also observed in Paper~\textsc{i} by analysing the ratios of [\ion{S}{ii}] and [\ion{N}{ii}] to H$\alpha$ and in agreement with measurements in nearby galaxies \citep{haffner09}.
In about half of the regions we observe zones of optically thin gas extending until the regions edge, which could potentially indicate escape channels cleared by stellar feedback. In Sect.~\ref{section:budget} we explore the link between the fraction of photons escaping the regions and the ionisation structure traced by [\ion{S}{ii}]/[\ion{O}{iii}].

In Table~\ref{table:obs_properties_regions} we summarise the visual appearance of the regions in [\ion{S}{ii}]/[\ion{O}{iii}], with a similar approach as \citet{pellegrini12}. Regions are classified
into a `classical' ionisation bounded nebula (`IB'), with an ionisation structure as the model in Fig.~\ref{fig:cloudy_model}, or a nebula featuring one (or more) optically thin channels (`CH') cutting through the low ionisation zone and reaching the regions edge.
We base our classification on the value of [\ion{S}{ii}]/[\ion{O}{iii}] along the region contour, as shown in Fig.~\ref{fig:hii_class}. In the Figure, the solid points indicate the median value of the ratio along the contour, and the grey bars span the entire range of observed values. In order to determine which regions provide a potential escape path, we consider [\ion{S}{ii}]/[\ion{O}{iii}] = 0.5 as fiducial limit for the transition to optically thick gas: regions having a ratio below this threshold along their contours are classified as CH; all other regions are classified as IB.
We test the effect of spatial binning on this classification by using both a finer and a coarser binning pattern, corresponding to S/N = 15 and 25 in H$\beta$. The classification remains unchanged, suggesting that spatial tessellation does not play a decisive role.
We stress that this classification does not take into account the geometry of the nebula and can only be conclusive when having a statistically significant sample of regions that allows to eliminate selection effects.

\begin{table*}
\centering
\caption{Spectral classification of WR stars.}
\begin{tabular}{ccccc}
\hline \hline
Obj. ID & Ra (J2000) & Dec (J2000) & Features/Reference & Classification \\ \hline
WR1 & 23:57:40.81 & -32:35:36.5 & (1) & WN2 \\
WR2 & 23:57:41.43 & -32:35:35.9 & (1) & WC4 \\  
WR3 & 23:57:45.74 &-32:34:37.4 & Nebular lines + \ion{He}{ii} 4686 are weak, H$\beta$ absorption & WN + O star \\
WR4 & 23:57:44.75 &-32:34:25.1 & \ion{N}{iii} 4640, \ion{He}{ii} 4886 & WN (late-type) \\
WR5 & 23:57:40.90 &-32:35:35.6 & \ion{He}{ii} 6560 (strong and broad) & WC4\\
WR6 & 23:57:40.53 & -32:35:33.1 & \ion{N}{iii} 4640, \ion{He}{ii} 4686 & WN (late-type) \\
WR7 & 23:57:41.37 &-32:35:52.5 & \ion{N}{v} 4603 + \ion{N}{iii} 4640 (in equal strength) & WN (mid-type) \\
WR8 & 23:57:43.27 & -32:35:49.6 & \ion{He}{ii} 4686 (narrow), \ion{N}{iii} 4640 & WN (late-type)\\
WR9 & 23:57:40.94 &-32:34:47.3  & Analogue to WR8, slightly broader \ion{He}{ii} & WN (late-type)\\
\hline
\end{tabular}
\tablebib{(1) \citet{bibby10}}
\label{table:wr_stars}
\end{table*}

\section{Ionisation budget}
\label{section:budget}

We construct a ionisation budget for the entire FoV and inspect the \ion{H}{ii} regions labelled in Fig.~\ref{fig:sii_oiii} in detail.
The latter regions were selected to lie for their most part in the FoV, and not at the edge of the field.
We note that we consider here the outermost contours of the \ion{H}{ii} regions sample, based on a flux brightness cut in H$\alpha$ = $\SI{6.7e-18}{\erg \per \second \per \centi \metre \squared}$spaxel$^{-1}$ as defined in Paper~\textsc{i}. We therefore consider the largest \ion{H}{ii} region complexes (such as regions 1 and 8 in Fig.~\ref{fig:sii_oiii}) in their entirety. This choice is made because dividing the regions within smaller subregions would imply arbitrary choices of their outermost border and division of their fluxes.

\par For each region, we model the expected $Q(H^0)$ from the stellar content as described below, and compare it to the observed ionising flux computed from the (reddening corrected) H$\alpha$ luminosity, following the calibration of \citet{kennicutt98}:
$$ Q(H^0)_{obs} = 7.31 \times 10^{11} L(H\alpha) \quad [\mbox{erg s}^{-1}],$$
based on the assumption of case B recombination and an electron temperature $T_e \sim \num{10 000}$ K.

\subsection{O stars}
Due to the selection method for the O stars catalogue (see Sect.~\ref{section:data}), we lack information such as mass, age or spectral class of the O stars. In order to compute what $Q(H^0)$ is expected from the O stars population, we therefore have to assume a mass distribution. We consider a Salpeter mass function \citep{salpeter55}:
$$ p(m) = A m^{-\alpha},$$
where $p(m)$ denotes the probability of finding a star of mass $m$, and $\alpha = 2.35$. We sample the distribution from $m_{min} = 20 M_\odot$, the lowest stellar mass probed by the catalogue, to $m_{max} = 60 M_\odot$ with 1000 Montecarlo realisations.
This upper limit corresponds to bright HII regions in the Milky Way \citep{feigelson13} and is the highest stellar mass tabulated in the work of \citet{martins05} (see below). We also obtain a lower limit for $Q(H^0)$ of field stars by sampling the Salpeter distribution up to a maximum mass  $m_{max} = 30~M_\odot$, corresponding to the most massive stars observed in typical galactic star forming regions \citep{bik10, gennaro12, bik12, ellerbroek13}. Moreover, stars of mass $M > 20~M_\odot$ are generally expected to emit enough ionising radiation to be embedded in an \ion{H}{ii} region, and if observed in the field are most likely runaway objects.
For each Montecarlo realisation, we compute the total number of O stars by taking into account the 30\% contamination rate and $\sim$ 70\% completeness of the stellar catalogue (see Sect.~\ref{section:data}):
$$n_{Ostars,tot} = \mathcal{N}(n_{Ostars}, 0.3),$$
where $\mathcal{N}$ denotes a normal distribution. Hereby, we also excluded O stars coinciding with the position of WR stars and YSCs (r $<$ 0.4'').
We then draw $n_{Ostars,tot}$ times a mass
$$m_i = p(m < m') = \int_{m_{min}}^{m'} p(m) dm = m_{min} \left[C \cdot \mbox{unif}(0,1) + 1 \right]^{1/(1 - \alpha)}$$
$$ C = \left( \frac{m_{max}}{m_{min}} \right)^{1 - \alpha} - 1,$$
where \texttt{unif} denotes a uniform distribution. The resulting masses are then converted into $Q(H^0)$ values by interpolation of Table 1 from \citet{martins05}, and the total flux of the region $Q(H^0)_{tot}$ is stored for each iteration.
We finally compute the best $Q(H^0)$ value and its uncertainty from, respectively, the median $m$ and $m \pm 1 \sigma$ of the resulting $Q(H^0)_{tot}$ distribution.

\subsection{YSC}
We determine $Q(H^0)$ for the YSC using the \textsc{slug} code, as described in Sect.~\ref{section:ysc_properties}.
We obtain a distribution of $Q(H^0)_{tot}$ from 1000 Montecarlo realisations: for each realisation, we sample one value from the PDF of each cluster in the region, and sum the resulting quantities. As for the O stars, we then compute the best $Q(H^0)$ value and its uncertainty from, respectively, the median $m$ and $m \pm 1 \sigma$ of the resulting $Q(H^0)_{tot}$ distribution.
Hereby, we also check that no YSCs are coincident with the position of WR stars, within a radius of 0.4''.

\subsection{WR stars}
Lastly, we compute the contribution of WR stars. We classify each star as Carbon-dominated (WC type) or \ion{He}{ii}-dominated (WN-type), based on its spectral features as listed in Table~\ref{table:wr_stars}. We then compute $Q(H^0)_{exp}$ from Tables 3 and 4 in \citet{smith02}, assuming a temperature $T = \num{60000}$ K for WN-type stars and $T = \num{120000}$ K for WC-type stars. We compute the uncertainty on $Q(H^0)$ with 1000 Montecarlo realisations of $T$ in the range $[40 - 80] \times 10^3$ K for WN-type stars and $[100 - 140] \times 10^3$ K for WC-type. Also in this case, we consider the median $m$ and $m \pm 1 \sigma$ of the resulting $Q(H^0)_{tot}$ distribution.

\begin{table*}
\centering
\caption{Observed properties of the sub-sample of regions shown in Fig.~\ref{fig:sii_oiii} and of the overall \ion{H}{ii} regions and DIG population. We note that the number of O stars indicated here is not completeness-corrected. $t_{max}$ indicates the age of the oldest cluster in each region (fiducial value of 50 $\pm$ 25 percentile of the age PDF; masking cluster with ages $> 10$ Myr). In the case of region 5, which is hosting exclusively O stars, we have assumed 3 Myr as an upper limit for the age. The indicated E(B-V) is the median extinction in the region (estimated from the Balmer decrement).
The total $L(H\alpha)$ in each region is reddening corrected. Regions are classified as described in Sect.~\ref{section:optical_depth} into `classical' ionisation bounded regions (IB) and regions featuring optically thin channels (CH).}
\begin{tabular}{lcccccccc}
\hline \hline
Region ID &
\multicolumn{3}{c}{Stellar content} &
$t_{max}$ &
$\log M_{GMC}$ & E(B-V) &
$\log L(H\alpha$) & Classification \\
 & \small{YSC} & \small{O$\star$} & \small{WR} & [Myr] & $[M_\odot]$ & & $[\mbox{erg s}^{-1}]$\\
\hline
\textbf{1} & 24 & 111 & 6 & $5.7_{-1.8}^{+1.8}$ & 6.1 & 0.14 & 39.11 & CH\\[0.1cm]
\textbf{2} & 3  & 30  & 1 & $4.3_{-2.9}^{+2.6}$ & 6.0 & 0.13 & 38.68 & CH\\[0.1cm]
\textbf{3} & 4  & 26  & 1 & $3.9_{-2.5}^{+2.9}$ & 5.3 & 0.19 & 38.73 & CH\\[0.1cm]
\textbf{4} & 1  & 19  & 0 & $5.7_{-1.4}^{+1.2}$ & 4.5 & 0.12 & 38.25 & IB\\[0.1cm]
\textbf{5} & 0  & 5   & 0 & $\leq 3.0$ & 3.4 & 0.14 & 37.35 & IB\\[0.1cm]
\textbf{6} & 1  & 2   & 0 & $1.7_{-1.2}^{+2.2}$ & --- & 0.09 & 37.41 & CH\\[0.1cm]
\textbf{7} & 1  & 2   & 0 & $3.2_{-2.1}^{+3.6}$ & --- & 0.11 & 37.01 & IB\\[0.1cm]
\textbf{8} & 8  & 110 & 1 & $4.3_{-2.6}^{+2.6}$ & 6.3 & 0.17 & 38.89 & CH\\[0.1cm]
\textbf{Tot \ion{H}{ii}} & 47 & 404 & 8 & & & & \\[0.1cm]
\textbf{Tot DIG} & 49 & 271 & 1 & & & &\\
\hline
\end{tabular}
\label{table:obs_properties_regions}
\end{table*}

\begin{table*}
\centering
\caption{Ionisation budget of the regions shown in Fig.~\ref{fig:sii_oiii} and of the overall \ion{H}{ii} regions and DIG population. The last row indicates a lower limit for the modelled ionising flux in the DIG, obtained by considering an upper limit $m_{max} = 30 M_\odot$ for the field candidate O stars (as opposed to $m_{max} = 60 M_\odot$ assumed elsewhere). All the uncertainties indicated are $\pm 1 \sigma$ errors on the $Q(H^0)$ distribution resulting from the Montecarlo sampling. The uncertainties on $\log Q^0_{obs}$ are all of order $\leq 0.001$.}
\begin{tabular}{lccccccc}
\hline \hline
Region ID &
$\log Q^0_{exp, Ostars}$ &
$\log Q^0_{exp, YSC}$ &
$\log Q^0_{exp, WR}$ &
$\log Q^0_{exp, tot}$ &
$\log Q^0_{obs}$&
$Q^0_{exp, tot}/Q^0_{obs}$ &
$f_{esc}$
\\
& [s$^{-1}$] & [s$^{-1}$]& [s$^{-1}$]& [s$^{-1}$] & [s$^{-1}$] & & \\
\hline

\textbf{1}
& $51.11_{-0.14}^{+0.14}$
& $49.60_{-0.34}^{+1.50}$
& $50.06_{-0.02}^{+0.02}$
& $51.16_{-0.13}^{+0.17}$
& $50.74$
& $2.71_{-0.78}^{+0.84}$
& $0.63_{-0.15}^{+0.09}$  \\[0.2cm]

\textbf{2}
& $50.54_{-0.13}^{+0.16}$
& $48.54_{-0.42}^{+1.59}$
& $49.40_{-0.09}^{+0.00}$
& $50.58_{-0.13}^{+0.16}$
& $50.31$
& $1.83_{-0.49}^{+0.65}$
& $0.45_{-0.20}^{+0.14}$  \\[0.2cm]

\textbf{3}
& $50.48_{-0.15}^{+0.17}$
& $48.54_{-0.43}^{+1.59}$
& $49.40_{-0.09}^{+0.00}$
& $50.52_{-0.15}^{+0.17}$
& $50.35$
& $1.50_{-0.47}^{+0.52}$
& $0.33_{-0.30}^{+0.17}$  \\[0.2cm]

\textbf{4}
& $50.34_{-0.15}^{+0.18}$
& $48.53_{-0.43}^{+0.59}$
& ---
& $50.34_{-0.15}^{+0.19}$
& $49.98$
& $2.31_{-0.78}^{+0.94}$
& $0.57_{-0.22}^{+0.13}$  \\[0.2cm]

\textbf{5}
& $49.70_{-0.25}^{+0.30}$
& ---
& ---
& $49.70_{-0.25}^{+0.30}$
& $49.10$
& $3.94_{-2.30}^{+2.74}$
& $0.75_{-0.36}^{+0.10}$  \\[0.2cm]

\textbf{6}
& $49.07_{-0.35}^{+0.94}$
& $48.53_{-0.43}^{+0.59}$
& ---
& $49.18_{-0.37}^{+0.86}$
& $49.14$
& $1.24_{-0.86}^{+1.81}$
& $0.20_{-1.83}^{+0.48}$  \\[0.2cm]

\textbf{7}
& $49.09_{-0.34}^{+0.86}$
& $48.53_{-0.43}^{+0.59}$
& ---
& $49.19_{-0.36}^{+0.80}$
& $48.75$
& $3.17_{-2.15}^{+4.09}$
& $0.68_{-0.67}^{+0.18}$  \\[0.2cm]

\textbf{8}
& $51.10_{-0.13}^{+0.14}$
& $48.85_{-0.42}^{+2.04}$
& $49.39_{-0.08}^{+0.01}$
& $51.11_{-0.13}^{+0.15}$
& $50.53$
& $3.85_{-1.15}^{+1.19}$
& $0.74_{-0.11}^{+0.06}$  \\[0.2cm]

\textbf{Tot \ion{H}{ii}}
& $51.68_{-0.13}^{+0.13}$
& $49.90_{-0.33}^{+1.04}$
& $50.23_{-0.02}^{+0.02}$
& $51.70_{-0.13}^{+0.14}$
& $51.22$
& $3.06_{-0.84}^{+0.93}$
& $0.67_{-0.12}^{+0.08}$  \\[0.2cm]

\textbf{Tot DIG}
& $51.50_{-0.12}^{+0.13}$
& $49.95_{-0.35}^{+0.91}$
& $49.40_{-0.09}^{+0.00}$
& $51.52_{-0.13}^{+0.15}$
& $50.65$
& $7.48_{-2.02}^{+2.39}$
& $0.87_{-0.05}^{+0.03}$\\[0.2cm]

\textbf{Tot DIG}$_{lower \thinspace limit}$
& $50.94_{-0.13}^{+0.13}$
&
&
& $50.99_{-0.15}^{+0.20}$
&
& $2.30_{-0.64}^{+0.75}$
& $0.57_{-0.17}^{+0.11}$\\
\hline
\end{tabular}
\label{table:photo_budget}
\end{table*}

\subsection{Resulting budget}
\label{section:budget_results}
Table~\ref{table:obs_properties_regions} and \ref{table:photo_budget} summarise, respectively, the observed properties and the photoionisation budget for the entire FoV
%(64 \ion{H}{ii} regions)
and in detail for the eight regions labelled in Fig.~\ref{fig:sii_oiii}.
In Table~\ref{table:obs_properties_regions} we list the stellar content, the age of the oldest star or cluster in each region, the total GMC mass, median $E(B - V)$ (estimated from the Balmer decrement) and H$\alpha$ luminosity (reddening corrected), as well as the morphological classification derived in Sect.~\ref{section:optical_depth}.
In Table~\ref{table:photo_budget} we indicate the total flux $Q(H^0)_{exp}$ modelled from the stellar content, as well as the relative contribution from O stars, YSC, and WR stars, and the observed ionising photon flux $Q(H^0)_{obs}$ derived from the reddening corrected H$\alpha$ luminosity. We also list the ratio of expected to observed flux and the corresponding escape fraction $f_{esc} = 1 - Q(H^0)_{obs}/Q(H^0)_{exp}$.
We note that the extremely large uncertainties on regions 5 and 7 are driven by the fact that these regions host (almost) exclusively O stars, for which $Q(H^0)$ is more loosely constrained. We also note that in region 6 the uncertainty ranges to unphysical values $f_{esc} < 0$, indicating that either the models are underestimating the photon flux, or the observed luminosity is being overestimated, possibly due to the reddening correction or to the exact location of the region boundaries.
 
\par From Table~\ref{table:photo_budget}, we see that WR and O stars dominate the contribution to the $Q^0_{exp, tot}$ value of the regions. Overall, we find an $f_{esc, \ion{H}{ii}} = 0.67_{-0.12}^{+0.08}$ for the entire population of \ion{H}{ii} regions (black contours in Fig.~\ref{fig:stellar_census}). We also observe that the stellar population in the DIG produces a more than sufficient amount of ionising photons ($Q(H^0)_{exp} > Q(H^0)_{obs}$), and that the DIG is therefore consistent with being self-ionised, with $f_{esc, DIG} = 0.87_{-0.05}^{+0.03}$. This holds also if considering a maximum mass of $30~M_\odot$ for field O stars (last row in Table~\ref{table:photo_budget}), in which case we find $f_{esc, DIG} = 0.57_{-0.17}^{+0.11}$.
In our FoV, we observe hence that the sources of ionising photons produce a photon flux that is more than sufficient to explain the emission of the ionised ISM, both within and outside the \ion{H}{ii} regions.

\subsection{Escape fraction from individual \ion{H}{ii} regions}
\label{section:fesc_individual}

In this section, we focus in more detail on individual \ion{H}{ii} regions. In order to better visualise the link between $f_{esc}$ and the visual appearance of the regions, in Fig.~\ref{fig:fesc_vs_appearance} we plot the ratio of $Q(H^0)_{exp}/Q(H^0)_{obs}$, and label each region according to the morphological criteria described in Sect.~\ref{section:optical_depth} (IB: ionisation bounded, CH: optically thin channels). We do not see any clear trend with visual appearance; we discuss this result in the light of other studies in Sect.~\ref{section:discussion}.

\par
\begin{figure*}%[ht]
%\center
\includegraphics[width=8cm]{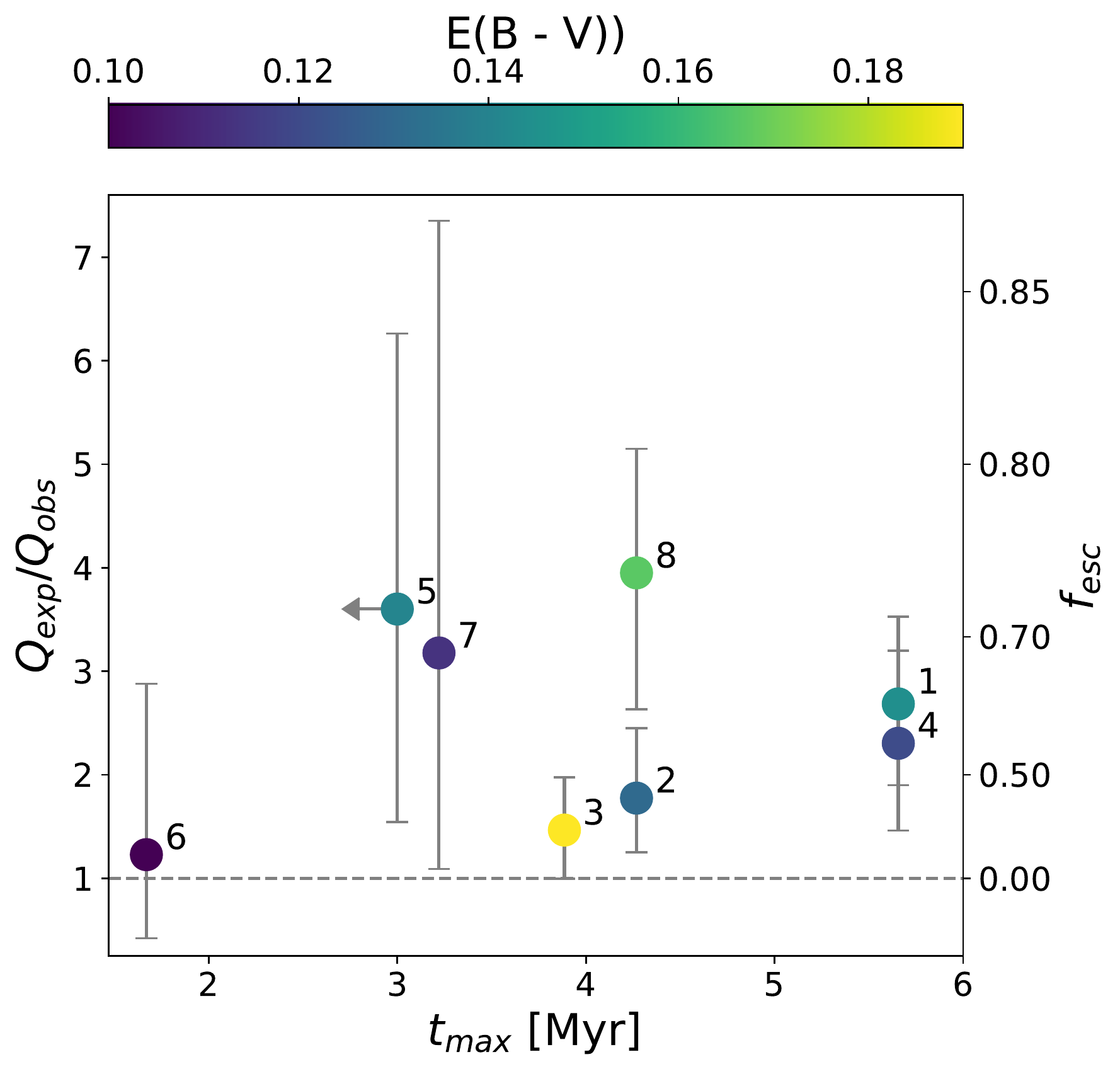}
\includegraphics[width=8cm]{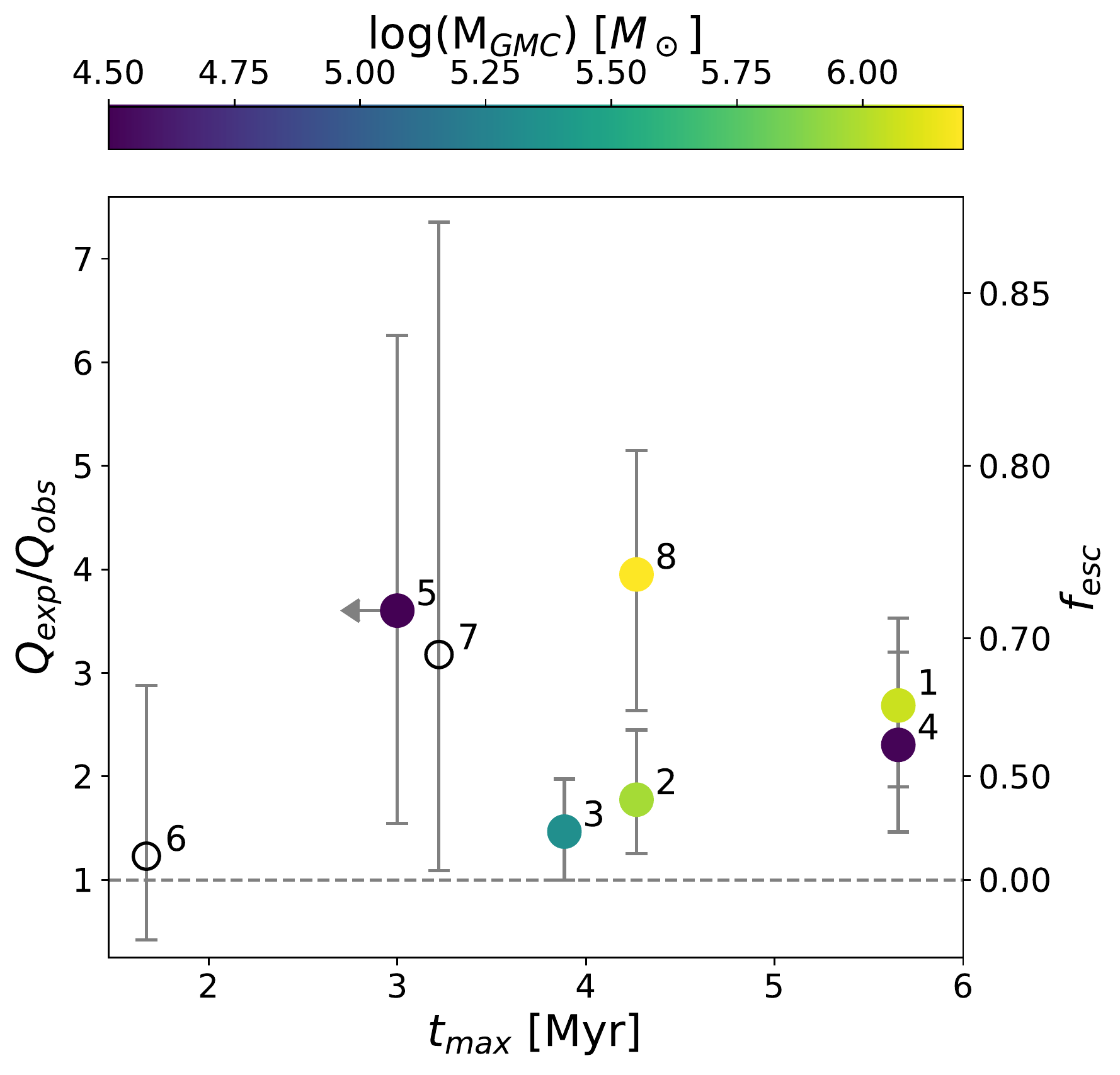}
\caption{Ratio of expected to observed ionising photon flux $Q(H^0)$ for the regions labelled in Fig.~\ref{fig:sii_oiii}, versus the age of the oldest cluster in each region (best-value estimate obtained from the median of the age PDF; masking cluster with ages $> 10$ Myr). For region 5, which is hosting exclusively O stars, we assume an upper limit of 3 Myr for the age. The dashed horizontal line indicates an $f_{esc} = 0$.
\textit{Left panel}: the points are colour-coded according to the median $E(B - V)$ in each region (Balmer decrement estimate). \textit{Right panel}: the regions are colour-coded according to their total GMC mass. Empty black circles indicate regions hosting no GMCs.}
\label{fig:fesc_vs_agerange}
\end{figure*}

 In Fig.~\ref{fig:fesc_vs_agerange}, we furthermore investigate the dependence of the $Q(H^0)_{exp}/Q(H^0)_{obs}$ ratio on the age of the oldest cluster in each region.
We consider the latter as a proxy for the age range spanned by the stellar population inside each \ion{H}{ii} region, assuming that all regions are currently forming stars.
We remark that this approximation might not be accurate for regions 6 and 7, in which no GMCs are detected with ALMA and the extinction traced by the ionised gas is very low (see Table~\ref{table:obs_properties_regions}); however, both regions are still hosting young candidate O stars.
When determining the age of the oldest cluster, we use as best estimate the median of each cluster PDF, and we exclude clusters with $t > 10$ Myr, as such objects are not anymore associated with \ion{H}{ii} regions, and are most likely line-of-sight objects. For region 5, which is hosting exclusively O stars, we assume an upper limit of 3 Myr for the age. 
We also colour-code the regions according to their median $E(B - V)$ (left panel) and total GMC mass (right panel), to indicate which regions are more affected by reddening and are still actively forming stars.
Dust can indeed absorb part of the LyC photons before they have the possibility to ionise hydrogen, and re-emit them at longer wavelengths, mimicking a larger $f_{esc}$.
We see that for regions 3 and 8, this could in part explain the $f_{esc} > 0$ observed, whereas - given their low dust and GMC content - regions 4 and 5 could be a true leaking region. We comment more on the effect of photons removal by dust in Sect.~\ref{section:discussion}. In our small sample of regions and due to the large uncertainties, we do not find any trend between $f_{esc}$ and the age of the stellar population in the region. This is discussed
in more detail in Sect.~\ref{section:discussion}.

\section{Discussion}
\label{section:discussion}
% Paper I
In Paper~\textsc{i}, we found a DIG fraction $f_{DIG, obs} = 0.15$ from the H$\alpha$ luminosity in our FoV\footnote{Revised estimate for the H$\alpha$-selected \ion{H}{ii} regions sample and corrected for extinction.}.
This is in good agreement with the fraction of diffuse gas expected by modelling the stellar and cluster population:
$$f_{DIG, exp} = \frac{Q^0_{exp,DIG} }{ Q^0_{exp,DIG}  + Q^0_{exp,HII}} \gtrsim 0.17.$$
Moreover, we assessed from our BPT diagram analysis in paper~\textsc{i} that the bulk of the DIG emission in our FoV was consistent with being photoionised. Here we confirm that the DIG is consistent with being self-ionised by field stars and clusters, with an overabundance of ionising photons $f_{esc} \sim 0.87_{-0.05}^{+0.03}$; we confirm this trend also when considering our lower-limit estimate.
This result is perhaps surprising considering the fact that we observe in Fig.~\ref{fig:sii_oiii} that the DIG is not optically thin to Lyman continuum emission (high [\ion{S}{ii}]/[\ion{O}{iii}] ratios). However, if the DIG has a clumpy structure, the emission line ratio could be tracing the denser, optically thick clumps; the apparent lack of H$\alpha$ emission can originate from photons escaping through the dilute, optically thin interclump medium, which through its low density will not contribute significantly to the observed emission line intensities.
We also observe that overall the \ion{H}{ii} regions are leaking ionising photons at a rate $f_{esc} \sim 0.67_{-0.12}^{+0.08}$, and that seven out of eight have an $f_{esc} \gtrsim 0.3$.

\par We note that the uncertainties on $f_{esc}$ are rather large, as we try to account for a variety of effects, including: the contamination rate $\sim 30\%$ in the O stars catalogue, the large range in $Q(H^0)$ obtained by Montecarlo sampling O stars of different mass,
the range in values spanned by the PDFs of the clusters and the assumption on the temperature of the WR stars.
The large uncertainties recovered are consistent with the work of \citet{niederhofer16}: by varying parameters of synthetic YSCs, the authors conclude that escape fractions derived from broadband photometry data are typically dominated by uncertainties in the spectral types of the stars.
However, even when considering the uncertainties, we find $f_{esc} > 0$ for the DIG overall and for six out of the eight \ion{H}{ii} regions inspected  (see Table~\ref{table:photo_budget} and Fig.~\ref{fig:fesc_vs_appearance}). 

\par An additional factor that is not taken into account in the computation of $f_{esc}$ is the effect of absorption of LyC photons by dust within the \ion{H}{ii} regions before these photons have the chance of ionising hydrogen atoms. In Sect.~\ref{section:extinction}, we have compared the extinction map derived from the H$\alpha$/H$\beta$ ratio with 24~$\mu$m emission from Spitzer/MIPS (Fig.~\ref{fig:extinction_map}, right panel), and found a good correspondence, indicative that dust and gas are well mixed \citep[see e.g.][]{choi20}. The resolution of mid-IR data is currently insufficient to confirm potential dust absorption in proximity of the sources of Lyman continuum photons at the distance of our target, and will only become possible with new generation IR telescopes such as JWST. However, in local \ion{H}{ii} regions, it is typically observed that a large fraction of LyC photons contributes to hydrogen ionisation. This fraction anti-correlates with metallicity \citep[e.g.][]{inoue01}, and at solar metallicity, an increasing fraction of Lyman continuum photons are absorbed by dust, instead. In the metallicity range spanned by our target (12 + log(O/H) $\sim  8.25 - 8.5$, see Sect.~\ref{section:metallicity}), \citet{inoue01} estimate that $\gtrsim 80\%$ Lyman continuum photons are absorbed by neutral hydrogen. Therefore, we expect that the effect of dust absorption in the \ion{H}{ii} regions in our FoV can reduce $Q(H^0)_{obs}$ of up to $\sim$ 20\%, which lies well within the given uncertainties.

% fesc 
\par Other studies targeting \ion{H}{ii} regions in nearby galaxies at high resolution found similar escape rates.
\citet{doran13} have estimated an $f_{esc} > 0.5$ in the 30 Dor star forming region in the LMC, based on a complete spectroscopic census of hot, luminous stars.
In a sample of \ion{H}{ii} regions in the LMC and SMC studied with narrowband photometry, \citet{pellegrini12} have found $f_{esc} > 0.4$ for all the regions. \citet{mcleod19} have studied in more detail two of the largest \ion{H}{ii} region complexes in the LMC with MUSE, finding $f_{esc} > 0.2$ for the complexes and the respective sub-regions.
Finally, in two regions recently observed with MUSE and HST in the nearby dwarf galaxy NGC 300, \citet{mcleod20} have estimated an $f_{esc} \gtrsim 0.3$ .

% DIG ionisation
Following the non-negligible escape of Lyman continuum photons from \ion{H}{ii} regions, several studies of local and nearby galaxies have proven the DIG consistent with being photoionised. For example, \citet{pellegrini12} have estimated that both in the LMC and SMC the total galactic escape fraction would be sufficient to account for the observed $f_{DIG}$, and that, when considering the additional contribution of field stars, the galaxies could leak a substantial amount of ionising radiation into the circumgalactic medium. Similarly, in a recent MUSE+HST study of the Antennae merger system, \citet{weilbacher18} have estimated a sufficient fraction of Lyman-continuum leakage in order to explain the amount of DIG observed.

% Optical depth
\par We do not find any correlation between ionising photon leakage in the HII regions and their visual appearance (`classic' ionisation bound regions vs regions featuring optically thin channels).
This is somewhat in contrast with the results of \citet{pellegrini12} that, by comparing the fiducial $f_{esc}$ assigned to \ion{H}{ii} regions in the LMC and SMC based on their appearance in [\ion{S}{ii}]/[\ion{O}{iii}] (see Sect.~\ref{section:optical_depth}) with previous $f_{esc}$ estimated based on the stellar population, found a good agreement overall. On the other hand, a systematic trend is not apparent from the MUSE results of \citet{mcleod19} on the LMC, that used the [\ion{O}{ii}]7320,7330/[\ion{O}{iii}]4959,5007
ratio as optical depth tracer. We interpret this apparent tension as likely being a simple effect of 3D geometry in number limited samples.
% Age of the stellar population
\par In our small sample of regions, we do not find evidence for a correlation between age range of the stellar population and $f_{esc}$. Increasing the sample size will be essential to investigate any such correlation, that could be expected in light of the results of galactic and cosmological simulations.
\citet{kim13a} simulated ionising radiation and supernova feedback in a low-redshift galactic disk, at a high spatial resolution $\sim 4$ pc. The study indicated that photons emitted at later ages are more likely to escape the star forming region, as the gas surrounding the region is dispersed by ionising radiation and supernova feedback. 
A similar result was obtained in the recent high-resolution cosmological simulations of \citet{ma20}, that studied $\sim$ 30 zoom-in simulations of galaxies at $z \gtrsim 5$.
The simulations suggest that ionising photons are preferably leaking from star forming regions containing a kpc-size superbubble, likely created by clustered SNe set off by stars of age $>$ 3 Myr. As the bubble expands, new stars are formed at its edge, inside a dense shell of compressed gas. The shell keeps expanding while forming stars and, as a consequence, the young stars formed in it end up inside the superbubble and are able to fully ionise channels of low-column density  pre-cleared by feedback from the previous population of stars. Therefore, regions hosting stars spanning a large range in age seem to be advantaged in leaking photons. Also observations have already pointed to the importance of the local star formation history in shaping the ionisation conditions in \ion{H}{ii} regions. By analysing a sample of $\sim$ 5000 \ion{H}{ii} regions from CALIFA \citep[Calar Alto Large Integral Field Area survey,][]{sanchez12}, \citet{sanchez15} found for example a correlation between the position occupied by the regions in BPT diagrams and the age and metallicity of their stellar population.

\par Finally, we comment on the impact of using different stellar population model in our \texttt{cluster\_slug} analysis on the cluster physical properties. The use of non rotating and single star models produces a lower limit to the total estimated $Q(H^0)$ and therefore to $f_{esc}$, in particular at ages larger than 3-4 Myr. Binaries and rotating stars have the same effect, that is they increase the production of ionising photons at ages older than 3 Myr with respect to predictions from non-rotating single star models \citep[see][]{leitherer14, gotberg19}.
Any existing trend between $f_{esc}$ and age range of the stellar population would therefore be reinforced by different model assumptions, as also reported in the simulations of \citet{ma20}.

\section{Conclusions}
\label{section:conclusions}
We have studied the ionised gas in the nearby galaxy NGC 7793 with MUSE, and complemented our observations with HST and ALMA data tracing the stellar content and the molecular gas.

\par We have constructed a census of YSCs, O stars, and WR stars in the MUSE FoV, and studied the properties of the stellar population. We modelled the age, mass, and ionising flux of YSCs with the stochastic stellar population synthesis code \textsc{slug}, finding that clusters located in \ion{H}{ii} regions have a higher probability to be younger, less massive, and to emit a higher number of ionising photons than clusters in the field.

\par We have investigated the link between the stellar population and the dense and ionised gas. We have contrasted the reddening map derived from individual stellar extinctions with the one constructed from the H$\alpha$/H$\beta$ ratio and found that they compare well, but that the latter correlates better with the position of GMCs traced by ALMA. We have estimated an oxygen abundance for the \ion{H}{ii} regions from the MUSE data, using the $S$-strong line method from \citet{pilyugin16}. We found a median abundance of $12 + \log(O/H) \sim 8.37$ with a scatter of 0.25 dex, in agreement with the previous estimate by \citet{pilyugin14}. The abundance map appears to be rich in substructures, especially surrounding YSCs and WR stars. We caution however against possible degeneracies with for example the temperature and density, and we do indeed observe a correlation with the ionisation state of the regions.

\par We have studied the ionisation structure of the \ion{H}{ii} regions using the [\ion{S}{ii}]6716,31/[\ion{O}{iii}]4959,5007 ratio as a proxy for the optical depth. We have focused on a subset of regions and classified them based on the value of [\ion{S}{ii}]/[\ion{O}{iii}] along their border into ionisation bounded or as featuring channels of optically thin gas. Finally, we have compiled a photoionisation budget for the entire FoV and for the subset of \ion{H}{ii} regions.

\par Overall, we find an escape fraction $f_{esc} = 0.67_{-0.12}^{+0.08}$ for the population of \ion{H}{ii} regions, and that the DIG in our FoV is more than consistent with being self-ionised, with an $f_{esc} = 0.87_{-0.05}^{+0.03}$. 
This holds even when considering a lower-limit estimate for the DIG flux, derived by assuming a maximum mass of $30~M_\odot$ for the field O stars. We furthermore find that the $f_{DIG, exp} \gtrsim 0.17$ obtained by modelling the DIG stellar population is in good agreement with the DIG fraction derived from the observed H$\alpha$ luminosity in Paper~\textsc{i}, $f_{DIG, obs} = 0.15$. 

\par We observe an $f_{esc} \gtrsim 0.3$ in seven out of the eight studied regions, and investigate how the $f_{esc}$ is linked to the regions properties. Among others, we do not find any trend between $f_{esc}$ and the visual appearance of the regions; we read this as the effect of 3D geometry in our number limited sample.

\begin{acknowledgements}
We would like to thank the anonymous referee for the constructive feedback on the second draft of the manuscript.
This work is based on observations collected at the European Southern Observatory under ESO programme 60.A-9188(A).
A.A. acknowledges the support of the Swedish Research Council, Vetenskapsr\aa{}det, and the Swedish National Space Agency (SNSA).
This project has received funding from the European Research Council (ERC) under the European Union's Horizon 2020 research and innovation programme (grant agreement No 757535).
This research made use of Astropy\footnote{http://www.astropy.org}, a community-developed core Python package for Astronomy \citep{astropy:2013, astropy:2018}.
\end{acknowledgements}
%\nocite{*}
\bibpunct{(}{)}{;}{a}{}{,} % to follow the A&A style
\bibliographystyle{aa}
\bibliography{paperII}

\end{document}